\documentclass[prl,twocolumn,superscriptaddress,nofootinbib]{revtex4-2}
\usepackage{graphicx}
\usepackage{amssymb}
\usepackage{bm}
\usepackage{natbib}
\usepackage{amsmath}
\usepackage{mathrsfs} 
\usepackage{amstext}
\usepackage[english]{babel}
\usepackage[table, svgnames, dvipsnames]{xcolor}
\usepackage[colorlinks=true,citecolor=Cerulean,linkcolor=RubineRed,urlcolor=Cerulean]{hyperref}
\newcommand{\be}{\begin{eqnarray}}
\newcommand{\ee}{\end{eqnarray}}
\newcommand{\no}{\nonumber}

\begin{document}
\title{Krylov Subspace Methods for Quantum Dynamics with Time-Dependent Generators}

\author{Kazutaka Takahashi\href{https://orcid.org/0000-0001-7321-2571} 
{\includegraphics[scale=0.05]{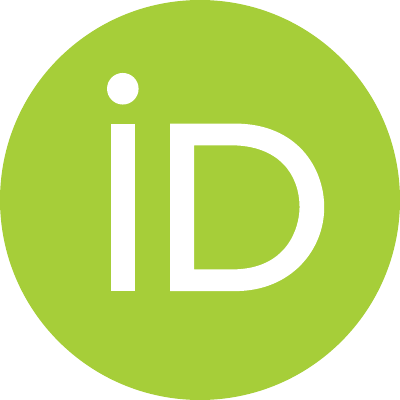}}}
\affiliation{Department  of  Physics  and  Materials  Science,  
University  of  Luxembourg,  L-1511  Luxembourg, Luxembourg}
\affiliation{Department of Physics Engineering, 
Faculty of Engineering, Mie University, Mie 514–8507, Japan}
\author{Adolfo del Campo\href{https://orcid.org/0000-0003-2219-2851}
{\includegraphics[scale=0.05]{orcidid.pdf}}}
\affiliation{Department  of  Physics  and  Materials  Science,  
University  of  Luxembourg,  L-1511  Luxembourg, Luxembourg}
\affiliation{Donostia International Physics Center,  E-20018 San Sebasti\'an, Spain}
\affiliation{Theoretical Division, Los Alamos National Laboratory, Los Alamos, New Mexico 87545, USA}

\begin{abstract} 
Krylov subspace methods in quantum dynamics identify the minimal subspace 
in which a process unfolds. 
To date, their use is restricted to time evolutions governed by 
time-independent generators. 
We introduce a generalization valid for driven quantum systems governed by 
a time-dependent Hamiltonian that maps the evolution to a diffusion problem 
in a one-dimensional lattice with nearest-neighbor hopping probabilities 
that are inhomogeneous and time dependent.  
This representation is used to establish a novel class of fundamental limits 
to the quantum speed of evolution and operator growth. 
We also discuss generalizations of the algorithm, adapted to discretized time evolutions 
and periodic Hamiltonians, with applications to many-body systems.
\end{abstract}

\maketitle

%%%%%%%%%%%%%%%%%%%%%%%%%%%%%%%%%%%%%%%%%%%%%%%%%%%%%%%%%%%%%%%%%%%%%%%%%%%%%%%
{\it Introduction}---The understanding of nonequilibrium quantum phenomena constitutes 
an open frontier of physics.  
Describing many-body quantum systems is challenging due to 
the large number of degrees of freedom involved. 
Approaches to reduce the complexity of the description are widely varied and 
involve approximation schemes such as 
effective theories and entanglement renormalization. 
Among them, Krylov subspace methods have long been recognized in applied mathematics 
for solving systems of linear algebraic equations~\cite{Liesen12} as well as 
in quantum physics to deal with systems with a large Hilbert space, 
ubiquitous in many-body problems~\cite{Viswanath94,nandy2024quantum}. 
Their use in the latter context has received a boost of attention, 
given recent applications to the study of quantum complexity and quantum 
chaos~\cite{Parker19,Barbon19,Rabinovici21,Caputa22,Balasubramanian22,Hornedal22}, 
quantum algorithms~\cite{Bharti21,Cortes22,Kirby23}, 
and quantum control~\cite{Takahashi24,Bhattacharjee23}. 
Krylov subspace methods have been developed for the evolution of 
operators~\cite{Parker19,Barbon19,Rabinovici21,Caputa22,Balasubramanian22,Hornedal22}, 
state vectors~\cite{Balasubramanian22}, density matrices~\cite{Caputa24}, and 
Wigner functions~\cite{Basu24}, as well as to tackle both unitary and 
open quantum dynamics~\cite{Bhattacharya22,LiuTangZhai23,Bhattacharya23,Bhattacharjee23}. 

Despite this progress, state-of-the-art techniques using 
Krylov subspace methods are restricted to time-independent generators. 
In general, treating nonautonomous equations 
with time-dependent generators is a difficult task.
In quantum dynamics, Hamiltonians do not commute at different times, 
and the evolution operator involves time ordering.
A naive extension of the method to time-dependent generators 
violates a key feature of the Krylov method, that is, 
the picture of the one-dimensional spreading.
We overcome this difficulty by introducing a Krylov algorithm
adapted to time-dependent generators that preserves 
the one-dimensional lattice representation of the dynamics. 
Using this general framework, we describe
near adiabatic dynamics for slow-varying systems~\cite{Born28,Kato50,Albash18}, 
the Krylov-space analog of shortcuts to adiabaticity for fast 
control~\cite{Demirplak03,Demirplak05,Berry09,Chen10,Muga10,Torrontegui13,GueryOdelin19}, 
and the Floquet theory for rapidly driven systems by a periodic 
field~\cite{Floquet1883,Shirley1965,Blanes09,Rudner20}.
We further introduce novel universal bounds on quantum dynamics for 
the characterization of operator growth and quantum evolution.

%%%%%%%%%%%%%%%%%%%%%%%%%%%%%%%%%%%%%%%%%%%%%%%%%%%%%%%%%%%%%
{\it Krylov construction for time-dependent generators}---We treat 
the Schr\"odinger equation, 
$i\partial_t|\psi(t)\rangle =\hat{H}(t)|\psi(t)\rangle$, 
governing the time evolution of the state vector 
$|\psi(t)\rangle$ under the time-dependent Hamiltonian $\hat{H}(t)$.
When the latter is time-independent, the state space is spanned by 
$\{\hat{H}^n|\psi(0)\rangle\}_{n=0}^\infty$, 
known as the Krylov space.
An orthonormal basis can be constructed by the standard Gram--Schmidt procedure.
This procedure brings the Hamiltonian to a tridiagonal form. 

An efficient way to explore the time-dependent case can be found from the so-called
$(t,t')$ formalism~\cite{Howland74,Peskin93,Peskin94,suppl}.
We introduce two different times, $s$ and $t$, to define    
\be
 |\psi(s,t)\rangle = e^{-is(\hat{H}(t)-i\partial_{t})}|\phi(t)\rangle, \label{tt'}
\ee
where $|\phi(t)\rangle$ is an arbitrary vector with the constraint
$|\phi(0)\rangle=|\psi(0)\rangle$.
The actual time-evolved state is obtained as $|\psi(t)\rangle=|\psi(t,t)\rangle$.
The operator $\hat{H}(t)-i\partial_t$ is sometimes used 
in the context of Floquet theory~\cite{Casati89,Grifoni98}.
This representation motivates us to use the space 
$\{(\hat{H}(t)-i\partial_t)^n|\phi(t)\rangle\}_{n=0}^\infty$.
If we consider extending the Hilbert space to a larger one,  
$\hat{H}(t)-i\partial_t$ is treated as a static 
Hamiltonian~\cite{Howland74,Peskin93,Peskin94,cao23,cao24}.
Here, we keep the space unchanged and 
the $(t,t')$ formalism is not exploited in the following discussions.

For a normalized basis $|K_0(t)\rangle$ with $|K_0(0)\rangle=|\psi(0)\rangle$,
we produce a series of time-dependent basis vectors by 
\be
 |K_{k+1}(t)\rangle b_{k+1}(t) =
 (\hat{H}(t)-i\partial_t)|K_k(t)\rangle \no\\
 -|K_k(t)\rangle a_k(t)-|K_{k-1}(t)\rangle b_k(t), \label{rec}
\ee
for $k=0,1,\dots,d-2$ with $|K_{-1}(t)\rangle b_0(t)=0$.
Each Lanczos coefficient is time dependent and is determined so that the orthonormal relation 
$\langle K_m(t)|K_n(t)\rangle=\delta_{m,n}$ holds.
We obtain 
\be
 && a_k(t)=\langle K_k(t)|(\hat{H}(t)-i\partial_t)|K_k(t)\rangle, \\
 && b_k(t)=\langle K_{k-1}(t)|(\hat{H}(t)-i\partial_t)|K_k(t)\rangle.
 \label{bk}
\ee
Here, each of $\{a_k(t)\}_{k=0}^{d-1}$ 
is real and the phase of $|K_k(t)\rangle$ 
is chosen so that each of $\{b_k(t)\}_{k=1}^{d-1}$ becomes non-negative.
The recurrence process halts when the possible basis vectors are exhausted, 
and the Krylov dimension $d$ is defined as the total number of the basis vectors.
The state vector is represented in the Krylov basis by the coherent quantum superposition  
$|\psi(t)\rangle=\sum_{k=0}^{d-1}|K_k(t)\rangle\varphi_k(t)$.
The transformed state  
$|\varphi(t)\rangle=\sum_{k=0}^{d-1}|k\rangle \varphi_k(t)
=(\varphi_0(t),\varphi_1(t),\dots,\varphi_{d-1}(t))^\mathrm{T}$,
satisfies $i\partial_t|\varphi(t)\rangle= \hat{\cal L}(t)|\varphi(t)\rangle$ 
with the initial condition $|\varphi(0)\rangle = |0\rangle$.
The generator $\hat{\cal L}(t)$ is written in a tridiagonal form as
$\hat{\cal L}(t)=\sum_{k=0}^{d-1}a_k(t)|k\rangle\langle k|
+\sum_{k=1}^{d-1}b_k(t)(|k\rangle\langle k-1|+|k-1\rangle\langle k|)$. 
Thus, the time evolution of the state can be described in terms of 
a single-particle hopping in a finite or semi-infinite chain, 
as in the time-independent case~\cite{Viswanath94,nandy2024quantum}.
The diagonal element $a_k(t)$ represents the on-site potential at site $k$ and 
the off diagonal $b_k(t)$ represents the hopping amplitude between $k-1$ and $k$.

The Krylov algorithm for the state 
involves diagonal components of the tridiagonal matrix.
They can be eliminated by the phase transformation
$|\tilde{K}_k(t)\rangle =|K_k(t)\rangle e^{-i\int_0^t ds\, a_k(s)}$. 
Then, the tridiagonal matrix has complex elements 
$\hat{\tilde{{\cal L}}}(t)=\sum_{k=1}^{d-1}(\tilde{b}_k^{\phantom *}(t)|k\rangle\langle k-1|
+\tilde{b}_k^*(t)|k-1\rangle\langle k|)$ 
with $\tilde{b}_k(t)=b_k(t)e^{i\int_0^t ds\, [a_k(s)-a_{k-1}(s)]}$.
A different possibility is to use the density operator 
$\hat{\rho}(t)=|\psi(t)\rangle\langle\psi(t)|$~\cite{suppl}.

The Krylov expansion is dependent on the choice of the initial basis. 
As in the time-independent case, 
it is natural to take $|K_0(t)\rangle=|\psi(0)\rangle$.
The first component $\varphi_0(t)=\langle\psi(0)|\psi(t)\rangle$ gives the survival amplitude.
An alternative choice for the time-dependent case is 
the instantaneous ground state of the Hamiltonian, $|\epsilon_0(t)\rangle$, 
under the condition that the initial state is in the ground state. 
Then, $|\varphi_0(t)|^2=|\langle \epsilon_0(t)|\psi(t)\rangle|^2$ represents 
the fidelity measuring how good the adiabatic approximation is.
The introduction of the time-dependent basis leads to the notion of parallel transport 
in Krylov space, associated with the generator $\hat{A}(t)$ satisfying 
$i\partial_t|K_k(t)\rangle=\hat{A}(t)|K_k(t)\rangle$. 
Our framework is then equivalent to the standard Krylov algorithm for 
$\hat{H}(t)-\hat{A}(t)$ defined locally in $t$.
The Hamiltonian is decomposed into two parts, which is reminiscent of shortcuts to 
adiabaticity~\cite{Demirplak03,Demirplak05,Berry09,Chen10,Muga10,Torrontegui13,GueryOdelin19}
where the counterdiabatic term, also known as adiabatic gauge potential, 
is used as a gauge field for parallel transport.

%%%%%%%%%%%%%%%%%%%%%%%%%%%%%%%%%%%%%%%%%%%%%%%%%%%%%%%%%%%%%%%%%%%%%%%
{\it Quantum speed and propagation limits in Krylov space}---A key advantage of 
the Krylov expansion is that, in any system,
it renders the state evolution as a propagation in a one-dimensional space. 
The degree of the propagation can be measured by 
the spread complexity~\cite{Parker19,Balasubramanian22}
\be
 K(t)=\sum_{k=0}^{d-1} k|\varphi_k(t)|^2.
\ee
It can be regarded as the expectation value of 
the Krylov operator $\hat{\cal K}=\sum_{k=0}^{d-1}k|k\rangle\langle k|$.
At early times of evolution, $K(t)$ increases from the initial value $K(0)=0$, 
and the growth rate is characterized by the Lanczos coefficients.
We can use the Robertson uncertainty relation to obtain 
the dispersion bound \cite{Hornedal22} as 
$|\partial_t K(t)|\le 2 \Delta L(t)\Delta K(t)$
where $\Delta L(t)=[\langle\varphi(t)|\hat{\cal L}^2(t)|\varphi(t)\rangle
-\langle\varphi(t)|\hat{\cal L}(t)|\varphi(t)\rangle^2]^{1/2}$
is the variance of $\hat{\cal L}(t)$ with respect to $|\varphi(t)\rangle$, 
and $\Delta K(t)$ is similarly defined for $\hat{\cal K}$. 
In contrast to the time-independent case \cite{Hornedal22},  
$\Delta L(t)$ is not equal to $b_1$.

Since we can define the Krylov operator, we can derive the operator 
quantum speed limit for the Heisenberg operator~\cite{Hornedal23}.
Here, we are interested in a speed limit for 
the time-evolved state rather than for the operator.
In the standard formulation, one considers as a distance 
the Fubini--Study angle $\Theta_\mathrm{FS}(t)
=\arccos|\langle\psi(0)|\psi(t)\rangle|$~\cite{Mandelstam45}. 
The minimum time for sweeping $\Theta_\mathrm{FS}(t)$ is then lower bounded by 
the time-average energy variance~\cite{Uhlmann92,Pfeifer93,delcampo21}.
To characterize the spreading, we generalize the Fubini--Study angle as 
\be
 \Theta_n(t) =\mathrm{arccos}\, \sqrt{P_n(t)}, \label{ep}
\ee
where $n=0,1,\dots,d-1$ and $P_{n}(t)=\sum_{k=0}^{n}|\varphi_k(t)|^2$ is the probability 
for the time-evolved state to remain within the first $n+1$ sites of the Krylov chain. 
We have generally 
$0=\Theta_{d-1}(t)\le\Theta_{d-2}(t)\le\dots\le \Theta_0(t)=\Theta_\mathrm{FS}(t)\le\pi/2$.
The spread complexity is written as $K(t)=\sum_{k=0}^{d-1}\sin^2\Theta_k(t)$.

By using similar techniques to those in the derivation of the celebrated Mandelstam--Tamm 
time-energy uncertainty relation~\cite{Mandelstam45,AnandanAharonov90}, 
we can derive upper bounds to $\Theta_n(t)$.
The time derivative of $\Theta_n=\Theta_n(t)$ leads to 
\be
 |\partial_t\Theta_n| \le b_{n+1}(t)
 \sqrt{1-\frac{\sin^2\Theta_{n+1}}{\sin^2\Theta_n}}
 \sqrt{1-\frac{\cos^2\Theta_{n-1}}{\cos^2\Theta_n}}, \label{qsl0}
\ee
where $n=0,1,\dots,d-2$ and $\cos^2\Theta_{-1}=0$ at $n=0$~\cite{suppl}.
The equality holds when the even components $\varphi_0(t), \varphi_2(t),\dots$ 
are real and the odd components $\varphi_1(t), \varphi_3(t),\dots$ are purely imaginary.
This is achieved when $\hat{\cal L}(t)$ has sublattice symmetry, that is, 
the diagonal components of $\hat{\cal L}(t)$ are equal to zero.
In the Krylov lattice, the condition $a_k(t)=0$ indicates a vanishing on-site potential. 
Then, no potential disturbs the wave function, and its spreading is maximized.

We can further simplify the right-hand side of Eq.~(\ref{qsl0}) by bounding it from above as 
$|\partial_t\Theta_n(t)|\le b_{n+1}(t)$.
This is a generalization of the Mandelstam--Tamm inequality
for the nonescape probability $P_{n}(t)$. 
Setting $n=0$ gives the Mandelstam--Tamm inequality
for the survival probability $|\langle K_0(t)|\psi(t)\rangle|^2$ 
adapted to any reference states~\cite{Suzuki20}.
If we choose $|K_0(t)\rangle$ properly,  
the upper bound $b_1(t)$ represents the energy variance with respect to $|K_0(t)\rangle$.

In the general case with $n\ne 0$, 
the bound cannot be tight except in the special case 
of a localized state $|\varphi_k(t)|\sim \delta_{k,n}$. 
By using $\sin\Theta_{n+1}(t)\ge 0$ and $\cos\Theta_n(t)\le 1$, 
we obtain tighter inequalities 
\be
 |\partial_t\Theta_n(t)|
 \le b_{n+1}(t)\sin\Theta_{n-1}(t)
 \le b_{n+1}(t)\Theta_{n-1}(t).
\ee
The last inequality can be iterated to obtain
\be
 \Theta_n(t)\le \int_0^{t} ds\,b_{n+1}(s)\int_0^{s} ds_n\, b_{n}(s_n)\dots
 \int_0^{s_2} ds_1 b_1(s_1).
 \no 
 \\ \label{lr}
\ee
When $b_k(t)$ takes a form $b_k(t)=b(t)c_k$, we can write 
\be
 \Theta_n(t)\le\exp\left[-(n+1)\ln\left(\frac{n+1}{v_{n+1}\tau(t)}\right)\right],
 \label{lr2}
\ee
for $n\gg 1$.
Here, $v_{n+1}=(c_1c_2\dots c_{n+1})^{1/(n+1)}$ represents the speed of propagation
and $\tau(t)=\int_0^tds\,b(s)$ represents an effective time scale.
The relation $v_n\tau(t)\sim n$ sets the scale of the propagation limit,
which is reminiscent of the Lieb--Robinson bound~\cite{Lieb72,Chen23}.
The derived inequalities are also applicable to systems 
with time-independent Hamiltonians.

%%%%%%%%%%%%%%%%%%%%%%%%%%%%%%%%%%%%%%%%%%%%%%%%%%%%%%%%%%%%%%%%%%%%%%%%%%%%%%%
{\it Examples with closed complexity algebra}---When the Hamiltonian has 
a tridiagonal form in a natural basis, and the initial state is 
chosen properly, the original basis is equivalent to the Krylov basis up to a phase. 
Paradigmatic examples are given by the following three systems~\cite{suppl}: 
(i) single spin $\hat{H}(t)=\bm{h}(t)\cdot\hat{\bm{S}}$, 
(ii) harmonic oscillator with translation 
$\hat{H}(t)=\frac{1}{2m}\hat{p}^2+\frac{m\omega^2}{2}[\hat{x}-x_0(t)]^2$, 
(iii) harmonic oscillator with dilation 
$\hat{H}(t)=\frac{1}{2m}\hat{p}^2+\frac{m}{2}\omega^2(t)\hat{x}^2$.
Also, when we choose the instantaneous ground state as the initial basis, 
each of the Krylov basis elements is given by an instantaneous energy eigenstate.
This is due to the property that the time-derivative operator 
in the instantaneous eigenstate basis is written in a tridiagonal form~\cite{suppl}.
The time derivative of the eigenstate is closely related to the counterdiabatic 
term~\cite{Demirplak03,Demirplak05,Berry09,Chen10,Muga10,Torrontegui13,GueryOdelin19}.
Generally, when the initial Krylov basis is chosen as an instantaneous eigenstate 
$|\epsilon_n(t)\rangle$, 
the first Lanczos coefficient $b_1(t)$ is given by the variance of the counterdiabatic term 
$\hat{H}_\mathrm{CD}(t)=i\sum_n(1-|\epsilon_n(t)\rangle\langle\epsilon_n(t)|)
|\partial_t \epsilon_n(t)\rangle\langle \epsilon_n(t)|$.

Remarkably, in each case (i)--(iii), 
the diagonal Lanczos coefficient $a_k(t)$ is linear in $k$ and 
the off diagonal part $b_k(t)$ takes a form $b_k(t) = b(t)c_k$, 
where the time-independent part $c_k$ is given by 
\be
 c_k = \left\{\begin{array}{ll}
 \sqrt{k(d-k)} & \mbox{for (i)},\\
 \sqrt{k} & \mbox{for (ii)},\\
 \sqrt{2k(2k-1)} & \mbox{for (iii)}.
 \end{array}\right.
\ee
The linear form of $a_k(t)$ reflects the property that 
these systems have constant eigenvalue spacing.
The time scale of the propagation limit determined from Eq.~(\ref{lr2}) is 
$\tau=O(n^{1/2})$ for (i) at $n\ll d$ and (ii), and $\tau=O(n^0)$ 
for (i) at $n\sim d$ and for (iii)~\cite{suppl}.

In the time-independent case, the forms of $c_k$ have been discussed as examples that give 
closed algebra of the Krylov operator $\hat{\cal K}$~\cite{Caputa22,Hornedal22}.
When we consider the phase-transformed basis $|\tilde{K}_k(t)\rangle$, 
we have $[\hat{\cal K},\hat{\tilde{{\cal L}}}(t)]=i\hat{\tilde{{\cal J}}}(t)$ and 
$[\hat{\tilde{{\cal J}}}(t),\hat{\cal K}]=i\hat{\tilde{{\cal L}}}(t)$,
where $\hat{\tilde{{\cal J}}}(t)=-i\sum_{k=1}^{d-1}(\tilde{b}_k^{\phantom *}(t)
|k\rangle\langle k-1|-\tilde{b}_k^*(t)|k-1\rangle\langle k|)$ 
is interpreted as the current operator.
The commutator of $\hat{\tilde{{\cal L}}}(t)$ and $\hat{\tilde{{\cal J}}}(t)$ generally gives
a diagonal matrix.
The algebra is closed for the above examples.
In each case, we find a form 
$[\hat{\tilde{{\cal L}}}(t),\hat{\tilde{{\cal J}}}(t)]=-i(\alpha(t)\hat{\cal K}+\gamma(t))$.
We have $\alpha\le 0$ for (i), $\alpha=0$ for (ii), and $\alpha\ge0$ for (iii).

In the time-independent case of these systems, the Heisenberg representation of $\hat{\cal K}$ is 
spanned by the identity operator, $\hat{\cal K}$, and $\hat{\tilde{{\cal J}}}$, and 
leads to the saturation of the operator quantum speed, 
maximizing the operator growth rate limit~\cite{Hornedal23}. 
For the time-dependent case, the evolving state is extended to a higher dimensional space 
and no saturation occurs \cite{suppl}.

%%%%%%%%%%%%%%%%%%%%%%%%%%%%%%%%%%%%%%%%%%%%%%%%%%%%%%%%%%
\begin{figure}[t]
\centering\includegraphics[width=1.\columnwidth]{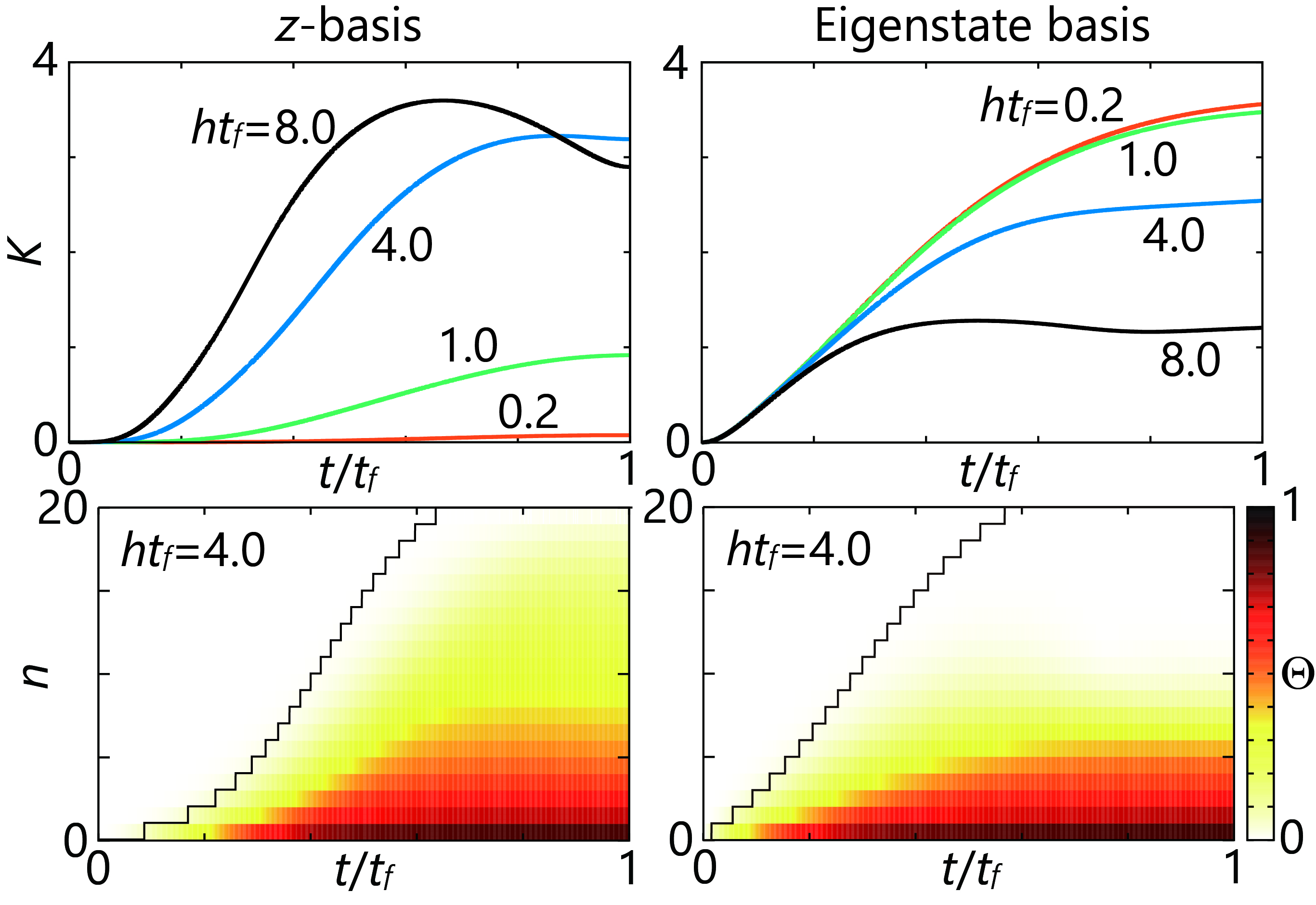}
\caption{Spreading in the Krylov space for the spin Hamiltonian
$\hat{H}(t)=h[\hat{S}^z\cos(\pi t/t_f)+\hat{S}^x\sin(\pi t/t_f)]$ 
with the spin quantum number $S=10$.
The left panels are for the fixed $z$ basis and 
the right panels are for the instantaneous eigenstate basis.
The upper panels represent the spread complexity $K(t)$ 
for several values of $ht_f$ and 
the lower panels represent $\Theta_n(t)$ at $ht_f=4.0$.
Each of the stepwise lines in the lower panels represents a boundary where 
the right-hand side of Eq.~(\ref{lr}) reaches a threshold value $\Theta_\mathrm{th}=0.1$.
}
\label{fig-spin}
\end{figure}
%%%%%%%%%%%%%%%%%%%%%%%%%%%%%%%%%%%%%%%%%%%%%%%%%%%%%%%%%%

We show in Fig.~\ref{fig-spin} the spread complexity $K(t)$ and $\Theta_n(t)$
for a single spin Hamiltonian.
We can consider two kinds of the Krylov basis: 
the fixed $z$-spin basis and the instantaneous eigenstate basis.
When we consider slow driving, the spread complexity grows rapidly in the fixed basis and is small in the instantaneous basis.
This behavior is reversed under a fast-driving scheme.
The spread complexity and $\Theta_n(t)$ in the instantaneous basis can be 
interpreted as a degree of nonadiabaticity.

%%%%%%%%%%%%%%%%%%%%%%%%%%%%%%%%%%%%%%%%%%%%%%%%%%%%%%%%%%%%%%%%%%%%%%%%%%%%%%%%%%%%%%%%%%
{\it Discrete time evolution and the Arnoldi iteration}---Discrete-time evolution 
arises in unitary quantum circuits. 
At the $k$th step, the quantum state $|\psi^k\rangle$ is propagated by the action of 
the unitary $\hat{U}^k$, as $|\psi^{k+1}\rangle=\hat{U}^k|\psi^k\rangle$.
For non-Hermitian matrices, the Krylov basis set is constructed by 
the Arnoldi iteration procedure~\cite{nandy2024quantum}.
Its applications range from open quantum systems~\cite{Bhattacharya23} to 
periodic systems~\cite{Yates21,Yates2022,Nizami23,Yeh24,Nizami24}, and 
unitary circuit dynamics~\cite{Suchsland23}.
The algorithm is adapted to the time-dependent case as 
\be
 && |K_{n+1}^{k+1}\rangle \sqrt{1-|z_n^k|^2}=\hat{U}^k|K_n^k\rangle-|f_n^{k+1}\rangle z_n^k, \\
 && |f_n^{k+1}\rangle = -|f_{n-1}^{k+1}\rangle\sqrt{1-|z_{n-1}^k|^2}
 +|K_n^{k+1}\rangle (z_{n-1}^k)^*,   
\ee
with $n=0,1,\dots,d-2$ and $z_{-1}^k=1$ at $n=0$.
The $n$th order Krylov basis $|K_n^k\rangle$ is defined at $n\le k$
and satisfies the orthonormal condition at each $k$.
We also use an auxiliary vector $|f_n^k\rangle$ and 
a complex number $z_n^k$ with $|z_n^k|\le 1$.
The operation $\hat{U}^k|K_n^k\rangle$ produces 
the one-step forwarded basis $|K_{n+1}^{k+1}\rangle$, instead of $|K_{n+1}^{k}\rangle$.
By expanding the time-evolved state as 
$|\psi^k\rangle =\sum_{n=0}^{\min (k,d-1)}|K_n^k\rangle\langle n|\varphi^k\rangle$,
we obtain the form $|\varphi^{k+1}\rangle=\hat{\cal U}^k|\varphi^k\rangle$
with $|\varphi^0\rangle=|0\rangle$.
The matrix $\hat{\cal U}^k$ takes an upper Hessenberg form.
Each component, Arnoldi coefficient, is written by using $z_n^k$~\cite{suppl}.

%%%%%%%%%%%%%%%%%%%%%%%%%%%%%%%%%%%%%%%%%%%%%%%%%%%%%%%%%%
\begin{figure}[t]
\centering\includegraphics[width=1.\columnwidth]{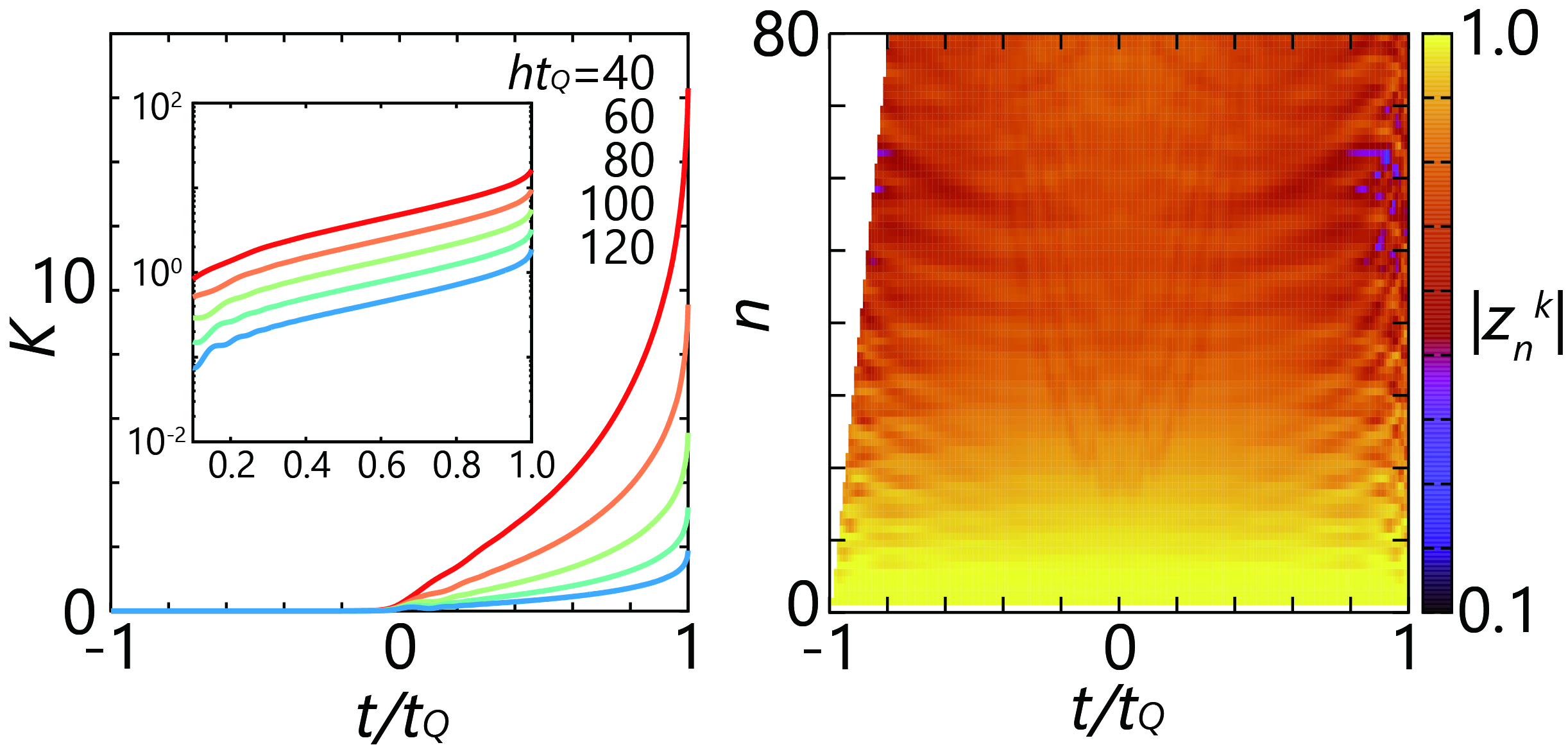}
\caption{Spreading in Krylov space for the one-dimensional quantum Ising model.
The left panel represents the spread complexity $K(t)$ for $N=24$ and $h\Delta t=0.1$, 
as functions of discrete time $t=-t_Q+k\Delta t$ with $-t_Q\le t\le t_Q$.
The inset in the left panel represents the log scale plot. 
The right panel represents $|z_n^k|$ with $n\le 80$ 
for $N=24$, $ht_Q=30$, and $h\Delta t=0.1$.
}
\label{fig-ising}
\end{figure}
%%%%%%%%%%%%%%%%%%%%%%%%%%%%%%%%%%%%%%%%%%%%%%%%%%%%%%%%%%

For illustration, we consider the one-dimensional quantum Ising model 
\be
 \hat{H}(t)=-\frac{h}{2}\left(1-\frac{t}{t_Q}\right)\sum_{i=1}^N \hat{X}_i
 -\frac{h}{2}\left(1+\frac{t}{t_Q}\right)\sum_{i=1}^N \hat{Z}_i\hat{Z}_{i+1}, 
 \no\\ 
 \label{ising}
\ee
where the operators on the right-hand side represent the Pauli operators~\cite{suppl}. 
Starting from the ground state of the Hamiltonian at $t=-t_Q$,
the system evolves until the final time $t=t_Q$.
The energy gap between the ground state and the first excited state of 
the Hamiltonian at $t=t_c=0$ goes to zero for $N\to\infty$ and 
the corresponding static system undergoes a quantum phase transition~\cite{Sachdev11}.

Figure~\ref{fig-ising} shows the growth of the spread complexity.
We take the instantaneous ground state as the initial basis.
For driving with a large quench time $t_Q$, the spread complexity is negligibly small 
at $t<t_c$ and starts growing around $t=t_c$ when the critical point is crossed. 
After the crossing, we observe an approximate exponential growth of $K(t)$ with respect to $t$ 
and an exponential suppression of $K(t=t_Q)$ with respect to $t_Q$. 
When the number of the iteration step $n$ is small, $|z_n^k|$ is close to unity, 
which means that the time evolution is described adiabatically.
Once the state spreads over $|K_n^k\rangle$ with higher $n$, 
small values of $|z_n^k|$ enhance nonadiabatic transitions.
Thus, we observe an effectively exponential growth of the spread complexity.

%%%%%%%%%%%%%%%%%%%%%%%%%%%%%%%%%%%%%%%%%%%%%%%%%%%%%%%%%%
\begin{figure}[t]
\centering\includegraphics[width=1.\columnwidth]{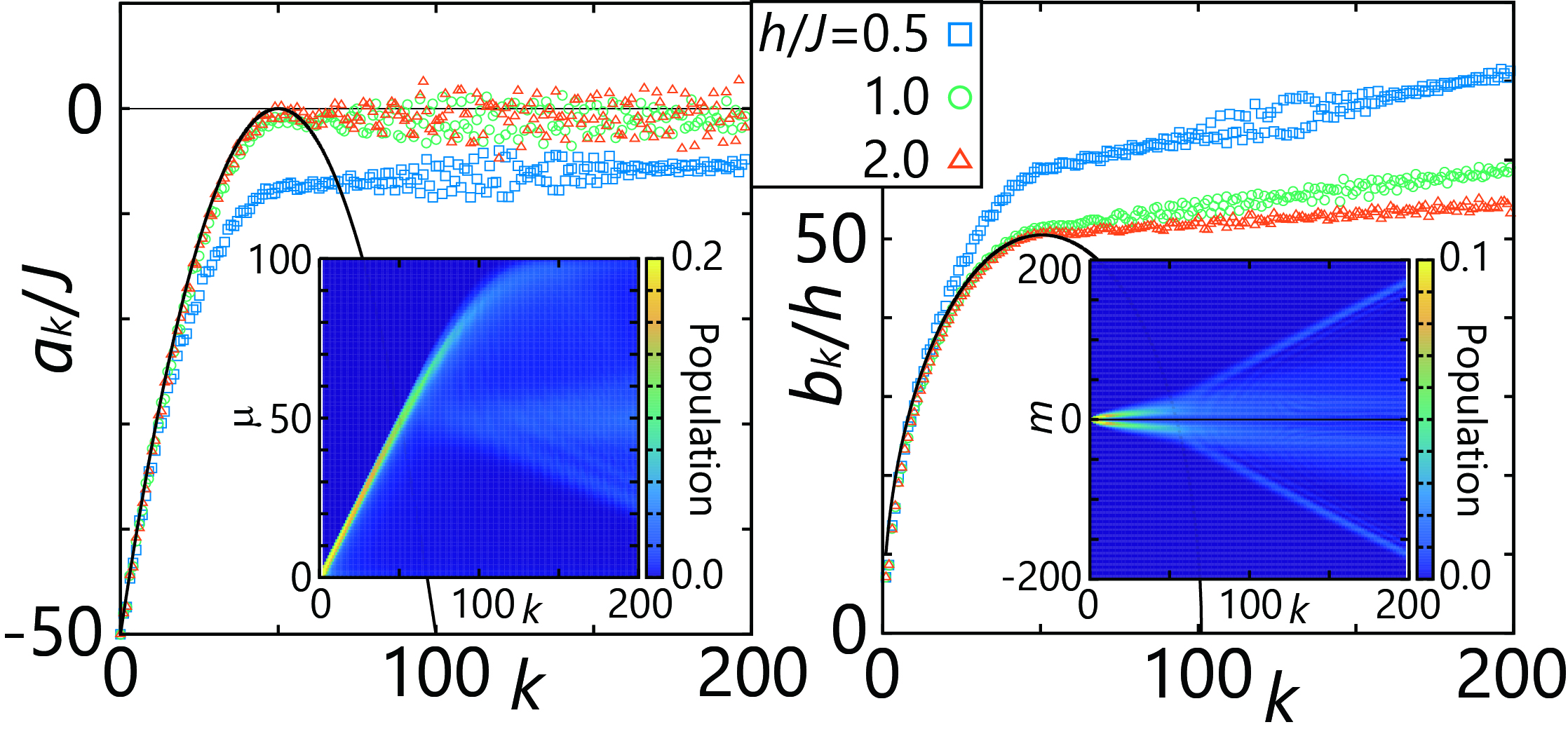}
\caption{The Lanczos coefficients for the Lipkin--Meshkov--Glick model.
We take $\Omega/J=0.1$ and $N=100$.
The solid lines represent the static limit $\Omega\to 0$.
The inset in the left panel is the population in the Hilbert space, and 
that in the right is the population in the Fourier space. 
We take $h/J=2.0$ in the insets.
}
\label{fig-lmg}
\end{figure}
%%%%%%%%%%%%%%%%%%%%%%%%%%%%%%%%%%%%%%%%%%%%%%%%%%%%%%%%%%

%%%%%%%%%%%%%%%%%%%%%%%%%%%%%%%%%%%%%%%%%%%%%%%%%%%%%%%%%%%%%%%%%%%%%%%%%%%%%%%%%%%%%%%%%%
{\it Periodically-driven systems}---When the Hamiltonian $\hat{H}(t)$ has 
a period $T=2\pi/\Omega$, 
we can apply the Floquet theorem~\cite{Floquet1883,Shirley1965}.
The Hilbert space is extended to the Floquet--Hilbert, or Sambe 
space~\cite{Sambe73}, which changes the definition of the inner product to  
include the time average over the period~\cite{suppl}.
The Krylov algorithm in Eq.~(\ref{rec}) is applied to give 
$|\psi(s,t)\rangle = \sum_{k=0}^\infty |K_k(t)\rangle\langle k|e^{-is\hat{\cal L}}|0\rangle$.
In this case, the tridiagonal matrix $\hat{\cal L}$ is time independent. 
The basis vectors are not orthogonal with each other in the usual sense, 
and the Krylov dimension is infinite even for systems in finite Hilbert space.
The time evolution by $\hat{\cal L}$ corresponds to that by the Floquet Hamiltonian, 
and that of the Krylov basis represents the micromotion.

We treat as an example the Lipkin--Meshkov--Glick model~\cite{LIPKIN1965}.
The Krylov analysis for the time-independent case has been studied  
for operators~\cite{Bhattacharjee22lmg} as well as for states~\cite{Afrasiar2023,Bento24}.
For the spin operator with $S=N/2$, we treat the Hamiltonian   
\be
 \hat{H}(t)=-\frac{2J}{N}(\hat{S}^z)^2-2h\hat{S}^x\sin\Omega t, \label{lmg}
\ee
with the initial condition $\hat{S}^z|\psi(0)\rangle=\frac{N}{2}|\psi(0)\rangle$.
We focus on a slow driving regime $\Omega\ll J$ where 
the standard Floquet picture is not applicable~\cite{WEINBERG2017}.
The static properties at $N\to\infty$ are described by the fixed-point analysis.
For $h<J$, $|\psi(0)\rangle$ corresponds to the stable ferromagnetic point.
It turns into an unstable point for $h>J$, and a stable paramagnetic point appears.

In Fig.~\ref{fig-lmg}, we show the Lanczos coefficients.
When the Krylov step $k$ is smaller than $N/2$, the half of the Hilbert space dimension, 
the Lanczos coefficients are close to those at the limit $\Omega\to 0$. 
To observe how the spreading occurs in the Krylov space, 
we numerically calculate 
the population in the $z$-spin eigenstates $|\mu\rangle$ with 
$\hat{S}^z|\mu\rangle = \left(\frac{N}{2}-\mu\right)|\mu\rangle$
as $\frac{1}{T}\int_0^T dt\,|\langle\mu|K_k(t)\rangle|^2$, 
and the population in each Fourier component 
$\left|\frac{1}{T}\int_0^T dt\,e^{im\Omega t}|K_k(t)\rangle\right|^2$, respectively. 
For $k<N/2$, the spreading is in the direction of the Hilbert space 
and the spreading in the Fourier space is suppressed.
When the step reaches the point $k=N/2$, 
$b_k$ at $\Omega\to 0$ decreases, 
which enhances the spreading in the Fourier space.
We find that $a_k$ is almost constant and $b_k$ shows a slow linear growth. 
We also find a qualitative difference between the results at $h<J$ and those at $h>J$.
For $h>J$ and $k<N/2$, the Lanczos coefficients rapidly converge to the static limit 
and are insensitive to the specific values of $h$.

%%%%%%%%%%%%%%%%%%%%%%%%%%%%%%%%%%%%%%%%%%%%%%%%%%%%%%%%%%%%%%%%%%%%%%%%%%%%%%%%%%%%%%%%%%
{\it Summary}---We have generalized the Krylov subspace
method for time evolutions of quantum states involving a time-dependent generator. Using it, the quantum state evolution is mapped to a one-dimensional hopping problem, as in the time-independent case. This description motivates generalized distance measures of spreading in Krylov space, for which we report universal bounds that are naturally related to quantum speed limits and Lieb--Robinson bounds.
We show how our method can be flexibly adapted to 
treat discrete-time evolution and periodic driving,
making it possible to analyze the many-body dynamics and quantum complexity of driven and controlled quantum processes. Our formulation for nonautonomous differential equations is rather general, and beyond quantum science, 
it should find broad applications 
in nonequilibrium phenomena, dynamical systems, and various fields of mathematics. 

{\it Acknowledgements}---We thank 
Budhaditya Bhattacharjee, Andr\'as Grabarits,  Aritra Kundu, Pratik Nandy, and Ruth Shir 
for stimulating discussions. 
A.~dC. thanks the Los Alamos National Laboratory for its hospitality during the completion of this work. 
We acknowledge the financial support 
from the Luxembourg National Research Fund (FNR Grant No. 16434093). 
This project has received funding from the QuantERA II Joint Programme with co-funding 
from the European Union’s Horizon 2020 research and innovation programme.
K.~T. further acknowledges support from JSPS KAKENHI Grant No. JP24K00547.

\bibliography{KrylovTDGen_lib}

%%%%%%%%%%%%%%%%%%%%%%%%%%%%%%%%%%%%%%%%%%%%%%%%%%%%%%%%%%%%%%%%%%%%%%%%%%%%%%%%%%%
%\appendix
\newpage
\onecolumngrid

\def\theequation{S\arabic{equation}}
\setcounter{equation}{0}

\begin{center}
\textbf{\large Supplemental Material}
\end{center}

%%%%%%%%%%%%%%%%%%%%%%%%%%%%%%%%%%%%%%%%%%%%%%%%%%%%%%%%%%%%%%%%%%%%%%%%%%%%%%%
\section{$(t,t')$ formalism and Krylov method}

In the $(t,t')$ formalism, we define a two-time dependent state as in Eq.~(\ref{tt'}). 
The operator $\hat{H}(t)-i\partial_t$ is written by the unitary time-evolution operator 
$\hat{U}(t)$ for the Hamiltonian $\hat{H}(t)$ as 
\be
 \hat{H}(t)-i\partial_t = -\hat{U}(t)i\partial_t\hat{U}^\dag(t).
\ee
Then, we can write for $|\phi(t)\rangle$ with 
$|\phi(0)\rangle=|\psi(0)\rangle$ as 
\be
 |\psi(s,t)\rangle= e^{-is(\hat{H}(t)-i\partial_t)}|\phi(t)\rangle
 =\hat{U}(t)e^{-s\partial_t}\hat{U}^\dag(t)|\phi(t)\rangle
 =\hat{U}(t)\hat{U}^\dag(t-s)|\phi(t-s)\rangle.
\ee
Setting $s=t$, we find $|\psi(t,t)\rangle=\hat{U}(t)|\psi(0)\rangle=|\psi(t)\rangle$.

A different way to show that the operator $e^{-is(\hat{H}(t)-i\partial_t)}$ is 
equivalent to the time evolution operator is to use the Lie--Trotter product formula as 
\be
 e^{-is(\hat{H}(t)-i\partial_t)}
 &=& \lim_{M\to\infty}\left(e^{-i\frac{s}{M}\hat{H}(t)}e^{-\frac{s}{M}\partial_t}\right)^M \no\\
 &=& \lim_{M\to\infty}
 e^{-i\frac{s}{M}\hat{H}(t)}e^{-i\frac{s}{M}\hat{H}(t-s/M)}
 e^{-i\frac{s}{M}\hat{H}(t-2s/M)}
 \cdots e^{-i\frac{s}{M}\hat{H}(t-(M-1)s/M)}e^{-s\partial_t} .
\ee
The last expression at $s=t$ takes the form of the time-ordered product, 
except the last factor.
The last factor brings $|\phi(t)\rangle$ back to $|\phi(0)\rangle=|\psi(0)\rangle$.
Thus, we can find the equivalence.

The two-time dependent state $|\psi(s,t)\rangle$ is more complicated than 
the original state $|\psi(t)\rangle$.
This can be understood by using an orthonormal complete set $\{|n\rangle\}_{n=0,1,\dots}$ as 
\be
 e^{-is(\hat{H}(t)-i\partial_t)}|0\rangle
 = \hat{U}(t)\hat{U}^\dag(t-s)|0\rangle =&\sum_n |n(t)\rangle\langle n(t-s)|0\rangle,
\ee
where $|n(t)\rangle=\hat{U}(t)|n\rangle$.
This state generally contains contributions from all of $n=0,1,\dots$ 
and is simplified to $|0(t)\rangle$ at $s=t$.

The time-evolution law of $|\psi(s,t)\rangle$ is written as 
\be
 i\partial_s|\psi(s,t)\rangle = (\hat{H}(t)-i\partial_t)|\psi(s,t)\rangle.
\ee
This equation can be solved if we can find the eigenstates and 
eigenvalues of $\hat{H}(t)-i\partial_t$. 
It is a formidable task and we can apply the Krylov algorithm to tridiagonalize the operator.
By using the Krylov basis introduced in Eq.~(\ref{rec}), we define 
\be
 \hat{F}(t)=\sum_{k=0}^{d-1}|K_k(t)\rangle\langle k|
 =(|K_0(t)\rangle, |K_1(t)\rangle, \dots, |K_{d-1}(t)\rangle).
\ee
When the dimension of the Hilbert space is denoted by $N$, 
$\hat{F}(t)$ is a $N\times d$ matrix and satisfies 
$\hat{F}^\dag(t)\hat{F}(t)=\hat{I}_d$, where $\hat{I}_d$ is 
the identity operator in the $d$ dimensional space.
We note that $\hat{F}(t)\hat{F}^\dag(t)$ is not necessarily equal to 
the identity operator in the Hilbert space.
Then, the definition of the basis vector leads to the relation 
\be
 (\hat{H}(t)-i\partial_t)\hat{F}(t)=\hat{F}(t)(\hat{\cal L}(t)-i\partial_t).
\ee
Expanding the state in the Krylov basis is equivalent to the transformation
\be
 |\psi(s,t)\rangle =\hat{F}(t)|\varphi(s,t)\rangle.
\ee
The transformed vector $|\varphi(s,t)\rangle$ has the size of the Krylov dimension.
It satisfies 
\be
 i\partial_s|\varphi(s,t)\rangle=(\hat{\cal L}(t)-i\partial_t)|\varphi(s,t)\rangle,
\ee
with the initial condition $|\varphi(0,t)\rangle=|0\rangle$.
The two-time state is written as 
\be
 |\psi(s,t)\rangle=\hat{F}(t)e^{-is(\hat{\cal L}(t)-i\partial_t)}|0\rangle 
 =\sum_{k=0}^{d-1} |K_k(t)\rangle\langle k|e^{-is(\hat{\cal L}(t)-i\partial_t)}|0\rangle.
\ee

One of the prominent features of the Krylov method is that 
some physical quantities are described only by the Lanczos coefficients.
A typical example that is known for the time-independent Hamiltonian 
is the survival amplitude~\cite{Parker19}.
This property still holds in the time-dependent case, as we see from the relation
$\langle K_0(t)|\psi(s,t)\rangle=\langle 0|e^{-is(\hat{\cal L}(t)-i\partial_t)}|0\rangle$.
The right-hand side is represented by the Lanczos coefficients and their derivatives.
We can find differential equations for the Lanczos coefficients 
if we know the overlap $\langle K_0(t)|\psi(s,t)\rangle$.

%%%%%%%%%%%%%%%%%%%%%%%%%%%%%%%%%%%%%%%%%%%%%%%%%%%%%%%%%%%%%%%%%%%%%%%%%%%%%%%
\section{Krylov method for the density operator}

In the standard applications of the Krylov subspace method to quantum dynamics, 
it is common to resort to the Heisenberg picture and 
describe the time evolution of an operator rather, 
as opposed to describing the time evolution of the quantum state.  
The Heisenberg representation of a quantum operator $\hat{X}$
evolving under a time-independent Hamiltonian $\hat{H}$ is given by 
$e^{i\hat{H}t}\hat{X}e^{-i\hat{H}t}$.
In the time-dependent case, we cannot apply the same method 
as the time-evolution operator requires time-ordering 
given the noncommutative property $[\hat{H}(t),\hat{H}(t')]\ne 0$.

However, treating the density operator $\hat{\rho}(t)$ is still possible.
The density operator satisfies the von Neumann equation, 
\be
 i\partial_t\hat{\rho}(t)=[\hat{H}(t),\hat{\rho}(t)].
\ee
We can use the recurrence relation 
\be
 b_{k+1}(t)\hat{O}_{k+1}(t) = (\hat{\cal L}_H(t)-i\partial_t)\hat{O}_k(t)-b_k(t)\hat{O}_{k-1}(t), 
\ee
where $\hat{\cal L}_H(t)(\cdot)=[\hat{H}(t),(\cdot)]$.
The Lanczos coefficients are given by 
$b_k(t)=(\hat{O}_{k-1}(t),(\hat{\cal L}_H(t)-i\partial_t)\hat{O}_{k}(t))$
with a properly-defined inner product $(\hat{A},\hat{B})$ for an arbitrary set of
operators $\hat{A}$ and $\hat{B}$.

When the density operator is expanded as 
\be
 \hat{\rho}(t)=\sum_{k=0}^{d-1}\varphi_k(t)\hat{O}_k(t),
\ee
the vector $|\varphi(t)\rangle=\sum_{k=0}^{d-1}|k\rangle \varphi_k(t)$
satisfies $i\partial_t|\varphi(t)\rangle=\hat{\cal L}(t)|\varphi(t)\rangle$ with 
\be
 \hat{\cal L}(t)=
 \sum_{k=1}^{d-1}b_k(t)(|k\rangle\langle k-1|+|k-1\rangle\langle k|)
 =\left(\begin{array}{cccccc}
 0 & b_1(t) &  0 & &&\\
 b_1(t) & 0  & b_2(t) &&&\\
 0 & b_2(t) & 0 &&&\\
 &&&\ddots&&\\
 &&&&0&b_{d-1}(t) \\
 &&&&b_{d-1}(t) & 0
 \end{array}\right).
\ee
The diagonal components are absent in the present case, and the coefficient
$b_k(t)$ is different from the corresponding one $b_k(t)$ for the state $|\psi(t)\rangle$.
We further note that $\hat{O}_k(t)$ is neither Hermitian nor positive definite.
Since the density operator satisfies the normalization condition
$\mathrm{Tr}\,\hat{\rho}(t)=1$, we have several additional conditions, such as 
$\sum_{k=0}^{d-1}\varphi_k(t)\mathrm{Tr}\,\hat{O}_k(t)=1$.

%%%%%%%%%%%%%%%%%%%%%%%%%%%%%%%%%%%%%%%%%%%%%%%%%%%%%%%%%%%%%%%%%%%%%%%%%%%%%%%
\section{Krylov speed limit inequality}

To derive Eq.~(\ref{qsl0}), we start by evaluating the time derivative of the angle 
defined in Eq.~(\ref{ep}), which yields the identity relation 
\be
 -2\partial_t\Theta_n(t)\sin\Theta_n(t)\cos\Theta_n(t)
 &=& -i\sum_{k=0}^n \left(
 \langle\varphi(t)|k\rangle\langle k|\hat{\cal L}(t)|\varphi(t)\rangle
 -\langle\varphi(t)|\hat{\cal L}(t)|k\rangle\langle k|\varphi(t)\rangle
 \right) \no\\
 &=& -ib_{n+1}(t)\left(\langle\varphi(t)|n\rangle\langle n+1|\varphi(t)\rangle
 -\langle\varphi(t)|n+1\rangle\langle n|\varphi(t)\rangle\right) \no\\
 &=& -2b_{n+1}(t)\mathrm{Im}\,\varphi_n(t)\varphi_{n+1}(t), 
\ee
and the inequality 
\be
 |\partial_t\Theta_n(t)|\sin\Theta_n(t)\cos\Theta_n(t)
 &\le& b_{n+1}(t)|\varphi_n(t)||\varphi_{n+1}(t)| \no\\
 &=& b_{n+1}(t)\sqrt{\sin^2\Theta_{n}(t)-\sin^2\Theta_{n+1}(t)}
 \sqrt{\cos^2\Theta_n(t)-\cos^2\Theta_{n-1}(t)}.
\ee
The equality holds when $\varphi_n(t)\varphi_{n+1}(t)$ is pure imaginary.
This condition is satisfied when $a_k(t)=0$. 
In that case, the equation for $\varphi_n(t)$ reads
\be
 i\partial_t \varphi_n(t)= b_{n}(t)\varphi_{n-1}(t)
 +b_{n+1}(t)\varphi_{n+1}(t).
\ee
This equation can be solved  with the initial condition 
$\varphi_n(0)=\delta_{n,0}$. The even components $\varphi_0(t), \varphi_2(t),\dots$ are real 
and the odd components $\varphi_1(t), \varphi_3(t),\dots$ are pure imaginary.
Thus, we obtain Eq.~(\ref{qsl0}).

%%%%%%%%%%%%%%%%%%%%%%%%%%%%%%%%%%%%%%%%%%%%%%%%cd%%%%%%%%%%%%%%%%%%%%%%%%%%%%%%%
\section{Solvable examples}

%%%%%%%%%%%%%%%%%%%%%%%%%%%%%%%%%%%%%%%%%%%%%%%%%%%%%%%%%%%%%%%%%%%%%%%%%%%%%%%
\subsection{Single spin}

We treat the single spin system, generally written as 
\be
 \hat{H}(t)=h(t)\bm{n}(t)\cdot\hat{\bm{S}}.
\ee
$\hat{\bm{S}}=(\hat{S}^x,\hat{S}^y,\hat{S}^z)$ 
represents the spin operator with $\hat{\bm{S}}^2=S(S+1)$.
The applied external field is characterized by the absolute value 
$h(t)$ and the direction
$\bm{n}(t)=(\sin\theta(t)\cos\varphi(t),\sin\theta(t)\sin\varphi(t),\cos\theta(t))$.
The Hilbert space is generally spanned by the eigenstates of $\hat{S}^z$ satisfying 
\be
 \hat{S}^z |m\rangle = m|m\rangle, 
\ee
with $m=-S,-S+1,\dots,S$.
We set $\hat{H}(0)=h(0)\hat{S}^z$ and the initial state as $|\psi(0)\rangle =|S\rangle$.

%%%%%%%%%%%%%%%%%%%%%%%%%%%%%%%%%%%%%%%%%%%%%%%%%%%%%%%%%%%%%%%%%%%%%%%%%%%%%%%
\subsubsection{Fixed $z$-basis}

In the fixed $z$-basis, the Hamiltonian is represented 
in a tridiagonal form, which means that we can naturally obtain a Krylov basis.
When the initial basis is set as $|K_0(t)\rangle=|S\rangle$, we obtain
\be
 (\hat{H}(t)-i\partial_t)|K_0(t)\rangle = \hat{H}(t)|S\rangle = Sh(t)\cos\theta(t) |S\rangle
 +\frac{1}{2}\sqrt{2S}h(t)\sin\theta(t) e^{i\varphi(t)}|S-1\rangle.
\ee
This gives 
\be
 && a_0(t)=Sh(t)\cos\theta(t), \\
 && b_1(t)=\frac{\sqrt{2S}}{2}h(t)\sin\theta(t), \\
 && |K_1(t)\rangle = e^{i\varphi(t)}|S-1\rangle.
\ee

In a similar way, we can calculate the higher-order contributions.
The point in the present example is that 
the Hamiltonian is written in terms of three kinds of operators and 
we can generally write its action in a given state $|m\rangle$ as 
\be
 \hat{H}(t)|m\rangle
 &=& h\left[S^z\cos\theta+\frac{1}{2}\left(
 S^+ e^{-i\varphi}+S^- e^{i\varphi}\right) \right]|m\rangle \no\\
 &=& hm\cos\theta |m\rangle
 +\frac{h}{2}e^{-i\varphi}\sqrt{(S-m)(S+m+1)}|m+1\rangle+\frac{h}{2}e^{i\varphi}
 \sqrt{(S+m)(S-m+1)}|m-1\rangle.
\ee
The state vector $|m\rangle$ is time-independent, and taking the time derivative 
to the state vector does not give any contribution.
As a result, we obtain
\be
 && |K_k(t)\rangle = e^{ik\varphi(t)}|S-k\rangle \qquad (k=0,1,\dots,2S), \\
 && a_k(t)=(S-k)h(t)\cos\theta(t)+k\dot{\varphi}(t) \qquad (k=0,1,\dots,2S), \\
 && b_k(t)=\frac{1}{2}h(t)\sin\theta(t)\sqrt{k(d-k)} \qquad (k=1,\dots,2S),
\ee
where the dot denotes the time derivative.
The Krylov dimension is equal to the Hilbert space dimension, i.e., $d=2S+1$.

%%%%%%%%%%%%%%%%%%%%%%%%%%%%%%%%%%%%%%%%%%%%%%%%%%%%%%%%%%%%%%%%%%%%%%%%%%%%%%%
\subsubsection{Instantaneous eigenstate basis}

The instantaneous eigenstates of the Hamiltonian are
given by the eigenstates of rotated-$\hat{S}^z$ satisfying 
\be
 \hat{R}(t)\hat{S}^z\hat{R}^\dag(t)|m(t)\rangle= m|m(t)\rangle, 
\ee
with $m=-S,-S+1,\dots,S$.
The rotation operator $\hat{R}(t)$ is written as 
\be
 \hat{R}(t)=\exp\left(-i\theta(t)\hat{\bm{S}}\cdot\bm{e}_\varphi(t)\right), 
\ee
with $\bm{e}_\varphi(t)=(-\sin\varphi(t),\cos\varphi(t),0)$.

We set the initial Krylov basis as 
\be
 |K_0(t)\rangle = |S(t)\rangle.
\ee
We apply $\hat{H}(t)-i\partial_t$ to this basis.
The action of the Hamiltonian only gives an eigenvalue, and 
we need to consider the time derivative of the instantaneous eigenstates.
The time derivative is closely related to the counterdiabatic 
term~\cite{Demirplak03,Demirplak05,Berry09,Torrontegui13,GueryOdelin19}.
Generally, for possible instantaneous eigenstates $|m(t)\rangle$, we can write
\be
 i\partial_t |m(t)\rangle = \hat{H}_\mathrm{CD}(t)|m(t)\rangle
 +|m(t)\rangle i\langle m(t)|\dot{m}(t)\rangle. \label{cd}
\ee
In the present case, the counterdiabatic term $\hat{H}_\mathrm{CD}(t)$ is given by~\cite{Berry09}
\be
 \hat{H}_\mathrm{CD}(t)=\bm{n}(t)\times\dot{\bm{n}}(t)\cdot\hat{\bm{S}},
\ee
and we have
\be
 i\partial_t|m(t)\rangle 
 &=& -\frac{i}{2}(\dot{\theta}(t)-i\dot{\varphi}(t)\sin\theta(t))e^{-i\varphi(t)}
 \sqrt{(S-m)(S+m+1)}|m+1(t)\rangle \no\\
 && +\frac{i}{2}(\dot{\theta}(t)+i\dot{\varphi}(t)\sin\theta(t))e^{i\varphi(t)}
 \sqrt{(S+m)(S-m+1)}|m-1(t)\rangle 
 -m\dot{\varphi}(t)(1-\cos\theta(t))|m(t)\rangle.
\ee
Comparing this to the recurrence relation of the Krylov basis, we obtain 
\be
 && |K_k(t)\rangle=(-i)^ke^{ik(\varphi(t)+\varphi_0(t))}|S-k(t)\rangle\qquad (k=0,1,\dots,2S), \\
 && a_k(t)=(S-k)[h(t)+\dot{\varphi}(t)(1-\cos\theta(t))]
 +k(\dot{\varphi}(t)+\dot{\phi}_0(t))\qquad (k=0,1,\dots,2S), \\
 && b_k(t)=\frac{1}{2}\sqrt{\dot{\theta}^2(t)+\dot{\varphi}^2(t)\sin^2\theta(t)}
 \sqrt{k(d-k)}\qquad (k=1,\dots,2S),
\ee
where $d=2S+1$ represents the Krylov dimension and 
\be
 e^{i\phi_0(t)}=\frac{\dot{\theta}(t)+i\dot{\varphi}(t)\sin\theta(t)}
 {\sqrt{\dot{\theta}^2(t)+\dot{\varphi}^2(t)\sin^2\theta(t)}}.
\ee

%%%%%%%%%%%%%%%%%%%%%%%%%%%%%%%%%%%%%%%%%%%%%%%%%%%%%%%%%%%%%%%%%%%%%%%%%%%%%%%
\subsection{Harmonic oscillator with translation}

We consider a one-dimensional single-particle in a harmonic oscillator potential.
The center of the potential is changed as a function of time 
keeping the potential shape invariant.
The Hamiltonian is given by 
\be
 \hat{H}(t)=\frac{1}{2m}\hat{p}^2+\frac{m\omega^2}{2}(\hat{x}-x_0(t))^2. \label{tr}
\ee
We set $x_0(0)=0$ and choose the initial state as the ground state of $\hat{H}(0)$.

%%%%%%%%%%%%%%%%%%%%%%%%%%%%%%%%%%%%%%%%%%%%%%%%%%%%%%%%%%%%%%%%%%%%%%%%%%%%%%%
\subsubsection{Fixed eigenstate basis}

The Hamiltonian is written as
\be
 \hat{H}(t) = \left(\hat{a}^\dag\hat{a}+\frac{1}{2}\right)\omega
 -\sqrt{\frac{m\omega}{2}}\omega x_0(t)(\hat{a}+\hat{a}^\dag)
 +\frac{m\omega^2}{2}x_0^2(t).
\ee
Here, we define the lowering operator
\be
 \hat{a}= \sqrt{\frac{m\omega}{2}}\hat{x}
 +\frac{i}{\sqrt{2m\omega}}\hat{p},
\ee
and the raising operator $\hat{a}^\dag$.
They give $\hat{a}|n\rangle=\sqrt{n}|n-1\rangle$ and $\hat{a}^\dag|n\rangle=\sqrt{n+1}|n+1\rangle$
for the number state $|n\rangle$ satisfying $\hat{a}^\dag\hat{a}|n\rangle=n|n\rangle$.

The number state basis $\{|n\rangle\}$ allows us to write 
the Hamiltonian in a tridiagonalized form.
When we set $|K_0\rangle = |0\rangle$, we obtain 
\be
 && |K_k(t)\rangle = \left(-\frac{\dot{x}_0(t)}{|\dot{x}_0(t)|}\right)^k |k\rangle 
 \qquad (k=0,1,\dots), \\
 && a_k(t)=\left(k+\frac{1}{2}\right)\omega+\frac{m\omega^2}{2}x_0^2(t) \qquad (k=0,1,\dots), \\
 && b_k(t)=\sqrt{\frac{m\omega}{2}}\omega |x_0(t)|\sqrt{k} \qquad (k=1,2,\dots).
\ee

%%%%%%%%%%%%%%%%%%%%%%%%%%%%%%%%%%%%%%%%%%%%%%%%%%%%%%%%%%%%%%%%%%%%%%%%%%%%%%%
\subsubsection{Instantaneous eigenstate basis}

The Hamiltonian can also be written with respect to 
the operator at each time defined as 
\be
 \hat{a}(t)= \sqrt{\frac{m\omega}{2}}(\hat{x}-x_0(t))+\frac{i}{\sqrt{2m\omega}}\hat{p}.
\ee
We obtain 
\be
 \hat{H}(t)=\left(\hat{a}^\dag(t)\hat{a}(t)+\frac{1}{2}\right)\omega,
\ee
and the instantaneous eigenstates are given by $|n(t)\rangle$ 
with $n=0,1,\dots$, where 
\be
 && \hat{a}(t)|n(t)\rangle=\sqrt{n}|n-1(t)\rangle, \\
 && \hat{a}^\dag(t)|n(t)\rangle=\sqrt{n+1}|n+1(t)\rangle.
\ee

The time derivative of the instantaneous eigenstate is written as in Eq.~(\ref{cd}).
The counterdiabatic term in the present case is given by~\cite{Muga10} 
\be
 \hat{H}_\mathrm{CD}(t)=\dot{x}_0(t)\hat{p}
 =-i \sqrt{\frac{m\omega}{2}}\dot{x}_0(t)(\hat{a}(t)-\hat{a}^\dag(t)).
\ee
This has a tridiagonal form in the instantaneous eigenstate basis.
When we set $|K_0(t)\rangle =|0(t)\rangle$, we obtain 
\be
 && |K_k(t)\rangle = \left(-i\frac{\dot{x}_0(t)}{|\dot{x}_0(t)|}\right)^k |k(t)\rangle 
 \qquad (k=0,1,\dots), \label{tr-k} \\
 && a_k(t)=\left(k+\frac{1}{2}\right)\omega \qquad (k=0,1,\dots), \label{tr-a} \\
 && b_k(t)=\sqrt{\frac{m\omega}{2}}|\dot{x}_0(t)|\sqrt{k} \qquad (k=1,2,\dots). \label{tr-b}
\ee

%%%%%%%%%%%%%%%%%%%%%%%%%%%%%%%%%%%%%%%%%%%%%%%%%%%%%%%%%%%%%%%%%%%%%%%%%%%%%%%
\subsection{Harmonic oscillator with dilation}

Let us again consider a Harmonic oscillator system.
We set $x_0(t)=0$ and make the frequency $\omega$ a time-dependent function.
The Hamiltonian is given by 
\be
 \hat{H}(t)=\frac{1}{2m}\hat{p}^2+\frac{m}{2}\omega^2(t)\hat{x}^2.
\ee
The initial state is set as the ground state of the initial Hamiltonian.

%%%%%%%%%%%%%%%%%%%%%%%%%%%%%%%%%%%%%%%%%%%%%%%%%%%%%%%%%%%%%%%%%%%%%%%%%%%%%%%
\subsubsection{Fixed eigenstate basis}

We write the Hamiltonian in terms of the annihilation operator at $t=0$:
\be
 \hat{a}= \sqrt{\frac{m\omega(0)}{2}}\hat{x}
 +\frac{i}{\sqrt{2m\omega(0)}}\hat{p}.
\ee
We obtain 
\be
 \hat{H}(t)=\frac{\omega^2(t)+\omega^2(0)}{2\omega(0)}
 \left(\hat{a}^\dag\hat{a}+\frac{1}{2}\right)
 +\frac{\omega^2(t)-\omega^2(0)}{4\omega(0)}
 \left((\hat{a}^\dag)^2+\hat{a}^2\right).
\ee
Since the last term lowers or raises the state by two,
the Krylov basis is given by $|2k\rangle$.
When we set $|K_0(t)\rangle = |0\rangle$, we obtain 
\be
 && |K_k(t)\rangle = \left(\frac{\omega^2(t)-\omega^2(0)}{|\omega^2(t)-\omega^2(0)|}\right)^k
 |2k\rangle \qquad (k=0,1,\dots), \\
 && a_k(t)=\left(2k+\frac{1}{2}\right)\frac{\omega^2(t)+\omega^2(0)}{2\omega(0)} 
 \qquad (k=0,1,\dots), \\
 && b_k(t)=\frac{|\omega^2(t)-\omega^2(0)|}{4\omega(0)}\sqrt{2k(2k-1)} \qquad (k=1,2,\dots).
\ee

%%%%%%%%%%%%%%%%%%%%%%%%%%%%%%%%%%%%%%%%%%%%%%%%%%%%%%%%%%%%%%%%%%%%%%%%%%%%%%%
\subsubsection{Instantaneous eigenstate basis}

The Hamiltonian is written with respect to the operator
\be
 \hat{a}(t)= \sqrt{\frac{m\omega(t)}{2}}\hat{x}
 +\frac{i}{\sqrt{2m\omega(t)}}\hat{p}.
\ee
The counterdiabatic term is given by~\cite{Muga10}
\be
 \hat{H}_\mathrm{CD}(t)=i\frac{\dot{\omega}(t)}{4\omega(t)}
 (\hat{a}^2(t)-(\hat{a}^\dag)^2 (t)).
\ee
When we set $|K_0(t)\rangle=|0(t)\rangle$, we obtain 
\be
 && |K_k(t)\rangle 
 = \left(i\frac{\dot{\omega}(t)}{|\dot{\omega}(t)|}\right)^k|2k(t)\rangle \qquad (k=0,1,\dots), \\
 && a_k(t)=\left(2k+\frac{1}{2}\right)\omega(t) \qquad (k=0,1,\dots), \\
 && b_k(t)=\frac{|\dot{\omega}(t)|}{4\omega(t)}\sqrt{2k(2k-1)} \qquad (k=1,2,\dots).
\ee

%%%%%%%%%%%%%%%%%%%%%%%%%%%%%%%%%%%%%%%%%%%%
\subsection{An example of speed limit}

As an exactly solvable model, we treat the Hamiltonian of the 
harmonic oscillator potential in Eq.~(\ref{tr}) with 
\be
 x_0(t)=vt. \label{vt}
\ee
As a Krylov basis set, we use the instantaneous set 
in Eqs.~(\ref{tr-k}), (\ref{tr-a}), and (\ref{tr-b}).
The Lanczos coefficients are independent of $t$ and we can find the exact form 
of $\varphi_k(t)$.

Since the Hamiltonian is written as 
\be
 \hat{H}(t)= e^{-ivt\hat{p}}\left(\frac{1}{2m}\hat{p}^2
 +\frac{m\omega^2}{2}\hat{x}^2\right)e^{ivt\hat{p}},
\ee
the equation for $|\psi(t)\rangle$ can be solved by moving to the rotating frame.
We can write 
\be
 |\psi(t)\rangle = e^{-ivt\hat{p}}e^{-i\hat{H}_vt}|0\rangle,
\ee
where 
\be
 \hat{H}_v
 = \frac{1}{2m}(\hat{p}-mv)^2+\frac{m\omega^2}{2}\hat{x}^2-\frac{mv^2}{2}
 = e^{-imv\hat{x}}\left(\frac{1}{2m}\hat{p}^2
 +\frac{m\omega^2}{2}\hat{x}^2-\frac{mv^2}{2}\right)e^{imv\hat{x}}.
\ee
When $v>0$, the Krylov basis is written as 
\be
 |K_k(t)\rangle = (-i)^k|k(t)\rangle= (-i)^ke^{-ivt\hat{p}}|k\rangle.
\ee
Then, the overlap is written as 
\be
 \varphi_k(t)=i^k e^{imv^2t/2}\sum_{n=0}^\infty
 e^{-i(n+1/2)\omega t}\langle k|e^{imv\hat{x}}|n\rangle\langle n|e^{-imv\hat{x}}|0\rangle.
 \label{sum}
\ee
The matrix element of $e^{imv\hat{x}}$ with respect to the eigenstate basis 
can be calculated by using the coherent state.
We obtain 
\be
 \langle k|e^{imv\hat{x}}|n\rangle=e^{-\gamma^2/2}\sum_{l=0}^{\min(n,k)}
 \frac{\sqrt{n!k!}}{l!(n-l)!(k-l)!}(i\gamma)^{n+k-2l},
\ee
where 
\be
 \gamma = \sqrt{\frac{mv^2}{2\omega}}.
\ee
We substitute this expression to Eq.~(\ref{sum}). 
The sums over $n$ and $l$ can be performed by inverting their order as 
\be
 \sum_{n=0}^\infty\sum_{l=0}^{\min(n,k)}(\dots)
 =\sum_{n=0}^k\sum_{l=0}^n (\dots)+\sum_{n=k+1}^\infty\sum_{l=0}^k(\dots)
 =\sum_{l=0}^k\sum_{n=l}^\infty(\dots).
\ee
We obtain
\be
 \varphi_k(t)=\frac{1}{\sqrt{k!}}\left[-\gamma(1-e^{-i\omega t})\right]^k
 e^{-\gamma^2 (1-e^{-i\omega t})}e^{i(\gamma^2-1/2)\omega t}.
\ee
The probability for the state $|\psi(t)\rangle$ at the $k$th order of the Krylov basis 
is given by 
\be
 |\varphi_k(t)|^2=\frac{1}{k!}\left(2\gamma\sin\frac{\omega t}{2}\right)^{2k}
 \exp\left[-\left(2\gamma\sin\frac{\omega t}{2}\right)^2\right]
 \sim
 \left(\frac{2\gamma\sin\frac{\omega t}{2}}{\sqrt{k}}\right)^{2k}
 \exp\left[-\left(2\gamma\sin\frac{\omega t}{2}\right)^2\right].
\ee
The last expression is applied for $k\gg 1$.
This allows us to estimate the propagation limit.
The time scale is set by
\be
 2\gamma\sin\frac{\omega t}{2}=\sqrt{k}. \label{limit1}
\ee

%%%%%%%%%%%%%%%%%%%%%%%%%%%%%%%%%%%%%%%%%%%%%%%%%%%%%%%%%%
\begin{figure}[t]
\centering\includegraphics[width=0.6\columnwidth]{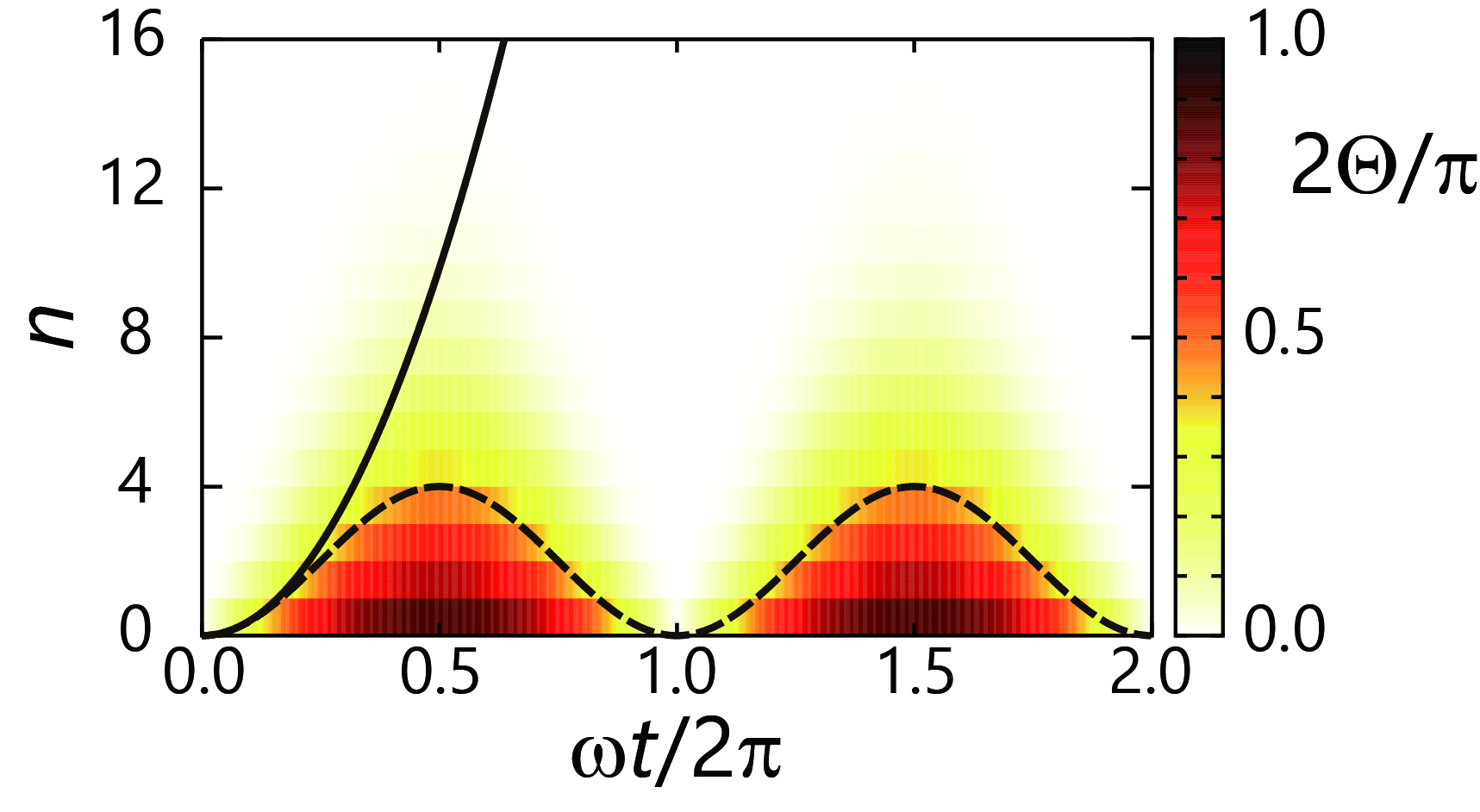}
\caption{
Spreading in the Krylov space for the harmonic-oscillator Hamiltonian 
in Eq.~(\ref{tr}) with Eq.~(\ref{vt}).
The solid line denotes a bound determined by Eq.~(\ref{limit2})
and the dashed line denotes Eq.~(\ref{limit1}).
}
\label{fig-ho}
\end{figure}
%%%%%%%%%%%%%%%%%%%%%%%%%%%%%%%%%%%%%%%%%%%%%%%%%%%%%%%%%%

On the other hand, $b_k(t)$ in the present case is given by 
\be
 b_k(t)=\omega\gamma \sqrt{k}.
\ee
The general propagation limit discussed in the main text is written as 
\be
 \Theta_n(t)\le\exp\left[-n
 \ln\left(\frac{\sqrt{n}}{\gamma\omega t}\right)\right], \label{limit}
\ee
for $n\gg 1$ and the time scale is identified as 
\be
 \gamma\omega t=\sqrt{n}. \label{limit2}
\ee
This is consistent with Eq.~(\ref{limit1}) at $\omega t\ll 1$.
The equality in Eq.~(\ref{limit}) holds only when $a_k(t)=0$.
In the present case, we have $a_k(t)=(k+1/2)\omega$.
Then, it is reasonable that the scale identified from Eq.~(\ref{limit})
coincides with Eq.~(\ref{limit1}) at $\omega t\ll 1$.

In Fig.~\ref{fig-ho}, we plot $\Theta_n(t)$.
Equation (\ref{limit2}) makes sense as a propagation limit at small $\omega t$.
Equation (\ref{limit1}) determined from the exact solution 
gives a good result in the whole range of $\omega t$.
\bigskip
%%%%%%%%%%%%%%%%%%%%%%%%%%%%%%%%%%%%%%%%%%%%%%%%%%%%%%%%%%%%%%%%%%%%%%%%%%%%%%%
\section{Complexity algebra}

%%%%%%%%%%%%%%%%%%%%%%%%%%%%%%%%%%%%%%%%%%%%
\subsection{Closed algebra}

For a given set of Lanczos coefficients, we can discuss 
the complexity algebra.
By using the phase transformation mentioned in the main body, 
we change the generator $\hat{\cal L}(t)$ as 
\be
 \hat{\tilde{{\cal L}}}(t) = \left(
 \begin{array}{cccc}
 0 & b_1(t) e^{i\delta_1(t)} & 0 &\\
 b_1(t) e^{-i\delta_1(t)} & 0 & b_2(t)e^{i\delta_2(t)} &\\
 0 & b_2(t)e^{-i\delta_2(t)} & 0 &\\
 &&& \ddots
 \end{array}\right),
\ee
where $\delta_k(t)=-\int_0^t ds\,(a_k(s)-a_{k-1}(s))$.
The corresponding current operator is given by 
\be
 \hat{\tilde{{\cal J}}}(t) = -i\left(
 \begin{array}{cccc}
 0 & -b_1(t)e^{i\delta_1(t)} & 0 &\\
 b_1 e^{-i\delta_1(t)} & 0 & -b_2(t)e^{i\delta_2(t)} &\\
 0 & b_2(t)e^{-i\delta_2(t)} & 0 &\\
 &&&\ddots
 \end{array}\right).
\ee
As we show in the main body, these operators are related to each other by
the commutator with the complexity operator 
\be
 \hat{\cal K}=\left(
 \begin{array}{cccc}
 0 & 0 & 0 &\\
 0 & 1 & 0 &\\
 0 & 0 & 2 &\\
 &&&\ddots
 \end{array}\right).
\ee
The spread complexity is written as $K(t)=\langle\varphi(t)|\hat{\cal K}|\varphi(t)\rangle$.
We have 
\be
 && [\hat{\tilde{{\cal K}}},\hat{\tilde{{\cal L}}}(t)]=i\hat{\tilde{{\cal J}}}(t), \label{klj}\\
 && [\hat{\tilde{{\cal J}}}(t),\hat{\cal K}]=i\hat{\tilde{{\cal L}}}(t).
\ee
These relations hold generally.
Furthermore, in the three systems studied in the previous sections, 
the commutator of $\hat{\tilde{{\cal L}}}$ and $\hat{\tilde{{\cal J}}}$ takes the form 
\be
 [\hat{\tilde{{\cal L}}}(t),\hat{\tilde{{\cal J}}}(t)]
 =-i(\alpha(t)\hat{\cal K}+\gamma(t)).
\ee
These commutation relations are not enough to describe the time evolution of 
the spread complexity as knowledge of the time dependence of the Lanczos coefficients 
is additionally required.
The equation is simplified when we impose 
\be
 && b_k(t)=b(t)c_k, \\
 && \delta_1(t) =\delta_2(t)=\delta_3(t)=\cdots=:\delta(t).
\ee
These conditions are satisfied in all the examples described in the previous section.
For example, for the single spin Hamiltonian in the instantaneous eigenstate basis, we can write  
\be
 && \alpha(t)=-\left(\dot{\theta}^2(t)+\dot{\varphi}^2(t)\sin^2\theta(t)\right), \\
 && \gamma(t)=-S\alpha(t), \\
 && b(t)=\frac{1}{2}\sqrt{-\alpha(t)}, \\
 && c_k=\sqrt{k(d-k)}, \\
 && \dot{\delta}(t)=
 h(t)-\dot{\varphi}(t)\cos\theta(t)-\dot{\phi}_0(t).
\ee
In the remaining part of the present section, we treat this case.

%%%%%%%%%%%%%%%%%%%%%%%%%%%%%%%%%%%%%%%%%%%%
\subsection{Time evolution of Heisenberg operators}

We consider the Heisenberg representation of three operators 
\be
 && \hat{\tilde{{\cal L}}}^\mathrm{H}(t)=\hat{\tilde{{\cal U}}}^\dag(t)
 \hat{\tilde{{\cal L}}}(t)\hat{\tilde{{\cal U}}}(t), \\
 && \hat{\tilde{{\cal J}}}^\mathrm{H}(t)=\hat{\tilde{{\cal U}}}^\dag(t)
 \hat{\tilde{{\cal J}}}(t)\hat{\tilde{{\cal U}}}(t), \\
 && \hat{\tilde{{\cal K}}}^\mathrm{H}(t)=\hat{\tilde{{\cal U}}}^\dag(t)
 \hat{\cal K}\hat{\tilde{{\cal U}}}(t), 
\ee
where $\hat{\tilde{{\cal U}}}(t)$ is the unitary time-evolution operator satisfying 
\be
 i\partial_t\hat{\tilde{{\cal U}}}(t)=\hat{\tilde{{\cal L}}}(t)\hat{\tilde{{\cal U}}}(t),
\ee
with $\hat{\tilde{{\cal U}}}(0)=1$.
The spread complexity is written as $K(t)=\langle\varphi(t)|\hat{\cal K}|\varphi(t)\rangle=
\langle 0|\hat{\tilde{{\cal K}}}^\mathrm{H}(t)|0\rangle$.
We apply the time derivative operator to these operators.
Due to the above-mentioned structure of the Lanczos coefficients, 
$\hat{\tilde{{\cal J}}}(t)$ is obtained from the time derivative of 
$\hat{\tilde{{\cal L}}}(t)$ and vice versa:
\be
 && \partial_t\hat{\tilde{{\cal L}}}(t)
 =\frac{\dot{b}(t)}{b(t)}\hat{\tilde{{\cal L}}}(t)
 +\dot{\delta}(t)\hat{\tilde{{\cal J}}}(t), \\
 && \partial_t\hat{\tilde{{\cal J}}}(t)
 =\frac{\dot{b}(t)}{b(t)}\hat{\tilde{{\cal J}}}(t)
 -\dot{\delta}(t)\hat{\tilde{{\cal L}}}(t).
\ee
We have
\be
 \partial_t\left(\begin{array}{c} \hat{\tilde{{\cal L}}}^\mathrm{H}(t) \\ 
 \hat{\tilde{{\cal J}}}^\mathrm{H}(t) \\ \hat{\tilde{{\cal K}}}^\mathrm{H}(t) 
 \end{array}\right)=\left(\begin{array}{ccc} \frac{\dot{b}(t)}{b(t)} & \dot{\delta}(t) & 0 \\
 -\dot{\delta}(t) & \frac{\dot{b}(t)}{b(t)} & \alpha(t) \\ 0 & 1 & 0 \end{array}\right)
 \left(\begin{array}{c} \hat{\tilde{{\cal L}}}^\mathrm{H}(t) \\ 
 \hat{\tilde{{\cal J}}}^\mathrm{H}(t) \\ \hat{\tilde{{\cal K}}}^\mathrm{H}(t) 
 \end{array}\right)+\gamma(t)\left(\begin{array}{c} 0 \\ 1 \\ 0 \end{array}\right).
 \label{ljk}
\ee
The current is written by the time derivative of the Krylov complexity as 
$\hat{\tilde{{\cal J}}}^\mathrm{H}(t)=\partial_t\hat{\tilde{{\cal K}}}^\mathrm{H}(t)$.
This relation comes from the commutation relation in Eq.~(\ref{klj}) and holds generally.

We consider the simplest case for the single spin system 
\be
\theta(t)=\omega t, \quad \varphi(t)=0, \quad \dot{h}(t)=0.
\label{ex}
\ee
The differential equation 
\be
 \partial_t\left(\begin{array}{c} \hat{\tilde{{\cal L}}}^\mathrm{H}(t) \\ 
 \hat{\tilde{{\cal J}}}^\mathrm{H}(t) \\ \hat{\tilde{{\cal K}}}^\mathrm{H}(t) 
 \end{array}\right)=\left(\begin{array}{ccc} 0 & h & 0 \\
 -h & 0 & -\omega^2 \\ 0 & 1 & 0 \end{array}\right)
 \left(\begin{array}{c} \hat{\tilde{{\cal L}}}^\mathrm{H}(t) \\ 
 \hat{\tilde{{\cal J}}}^\mathrm{H}(t) \\ \hat{\tilde{{\cal K}}}^\mathrm{H}(t) 
 \end{array}\right)+S\omega^2\left(\begin{array}{c} 0 \\ 1 \\ 0 \end{array}\right),
\ee
is solved analytically as 
\be
 \hat{\tilde{{\cal K}}}^\mathrm{H}(t) &=&  
 -\frac{h\left[1-\cos\left(\sqrt{h^2+\omega^2}t\right)\right]}{h^2+\omega^2}\hat{\cal L}(0)+
 \frac{\sin\left(\sqrt{h^2+\omega^2}t\right)}{\sqrt{h^2+\omega^2}}\hat{\cal J}(0) \no\\
 && +\frac{h^2+\omega^2\cos\left(\sqrt{h^2+\omega^2}t\right)}{h^2+\omega^2}\hat{\cal K}
 +S\frac{\omega^2\left[1-\cos\left(\sqrt{h^2+\omega^2}t\right)\right]}{h^2+\omega^2},
\ee
and 
\be
 && \hat{\tilde{{\cal J}}}^\mathrm{H}(t)=\partial_t\hat{\tilde{{\cal K}}}^\mathrm{H}(t), \\
 && \hat{\tilde{{\cal L}}}^\mathrm{H}(t)= h(\hat{\tilde{{\cal K}}}^\mathrm{H}(t)-\hat{\cal K})+
 \hat{\cal L}(0).
\ee
The spread complexity is given by
\be
 K(t)= S\frac{\omega^2\left[1-\cos\left(\sqrt{h^2+\omega^2}t\right)\right]}{h^2+\omega^2}.
\ee

%%%%%%%%%%%%%%%%%%%%%%%%%%%%%%%%%%%%%%%%%%%%
\subsection{Operator quantum speed limit}

In \cite{Hornedal23}, 
the operator quantum speed limit was derived.
When the formula is applied to the complexity operator $\hat{\cal K}$, we can write 
\be
 \mathrm{arccos}\,\left(\frac{\mathrm{Tr}\,\hat{\tilde{{\cal K}}}\hat{\tilde{{\cal U}}}^\dag(t)
 \hat{\tilde{{\cal K}}}\hat{\tilde{{\cal U}}}(t)}
 {\mathrm{Tr}\,\hat{\tilde{{\cal K}}}^2}\right) \le \int_0^t ds
 \sqrt{\frac{\mathrm{Tr}\,\hat{\tilde{{\cal J}}}^2(s)}{\mathrm{Tr}\,\hat{\tilde{{\cal K}}}^2}}, 
 \label{oqsl}
\ee
where  
\be
 \hat{\tilde{{\cal K}}}=\hat{\cal K}-\frac{1}{d}\mathrm{Tr}\,\hat{\cal K}.
\ee
This speed limit relation makes sense when $d$ is finite.
In the single spin case, we can write 
\be
 \mathrm{arccos}\,\left(\frac{\mathrm{Tr}\,\hat{\tilde{{\cal K}}}\hat{\tilde{{\cal U}}}^\dag(t)
 \hat{\tilde{{\cal K}}}\hat{\tilde{{\cal U}}}(t)}
 {\mathrm{Tr}\,\hat{\tilde{{\cal K}}}^2}\right)
 \le \int_0^t ds\,\sqrt{\dot{\theta}^2(s)+\dot{\varphi}^2(s)\sin^2\theta(s)}.
\ee
In the time-dependent case, the equality condition is not satisfied.
For example, when we use Eq.~(\ref{ex}), we obtain
\be
 \mathrm{arccos}\,\left(\frac{h^2+\omega^2\cos(\sqrt{h^2+\omega^2}t)}{h^2+\omega^2}\right)
 \le \omega t.
\ee
The equality is satisfied only in the trivial case $h=0$.

%%%%%%%%%%%%%%%%%%%%%%%%%%%%%%%%%%%%%%%%%%%%%%%%%%%%%%%%%%%%%%%%%%%%%%%%%%%%%%%
\section{Krylov method for discrete-time systems}

%%%%%%%%%%%%%%%%%%%%%%%%%%%%%%%%%%%%%%%%%%%%%%%%%%%%%%%%%%%%%%%%%%%%%%%%%%%%%%%
\subsection{Time evolution operator}

In the discrete-time case, the Krylov basis is produced by the Arnoldi
iteration procedure.
Then, by using the Arnoldi coefficients, 
we can consider the time evolution of the transformed state as 
$|\varphi^{k}\rangle =\hat{\cal U}^{k-1}|\varphi^{k-1}\rangle$, 
with $|\varphi^0\rangle=|0\rangle$.
The Arnoldi procedure is shown in the main body.
Here, we show how to construct $\hat{\cal U}^k$ in the iteration procedure.

We decompose the operator as
\be
 \hat{\cal U}^{k-1}=
 \sum_{l=0}^{\mathrm{min}\,(k-1,d-1)}|u_l^{k-1}\rangle\langle l|,
\ee
to write 
\be
 |\varphi^{k}\rangle=\hat{\cal U}^{k-1}|\varphi^{k-1}\rangle
 =\sum_{l=0}^{\mathrm{min}\,(k-1,d-1)}|u_l^{k-1}\rangle \varphi_l^{k-1}.
\ee
We produce the first vector from the Arnoldi procedure, as 
\be
 |u_0^{k-1}\rangle = |0\rangle z_0^{k-1}+|1\rangle
 \sqrt{1-|z_0^{k-1}|^2}.
\ee
Then, a perpendicular vector is defined as an auxiliary vector:
\be
 |v_0^{k-1}\rangle = -|0\rangle\sqrt{1-|z_0^{k-1}|^2}+|1\rangle (z_0^{k-1})^*.
\ee
We repeat the same procedure with  
\be
 && |u_l^{k-1}\rangle = |v_{l-1}^{k-1}\rangle z_l^{k-1}
 +|l+1\rangle\sqrt{1-|z_l^{k-1}|^2}, \\
 && |v_l^{k-1}\rangle = -|v_{l-1}^{k-1}\rangle \sqrt{1-|z_l^{k-1}|^2}
 +|l+1\rangle(z_l^{k-1})^*,
\ee
for $l=1,2,\dots,d-2$.
Finally, for $l=d-1$, we use 
\be
 |u_{d-1}^{k-1}\rangle = |v_{d-2}^{k-1}\rangle.
\ee
By completing this procedure, we can obtain 
all components of $\hat{\cal U}^{k-1}$.

The first three vectors are written explicitly as 
\be
 \hat{\cal U}=\left(\begin{array}{cccc}
 z_0 & -\sqrt{1-|z_0|^2}z_1 & \sqrt{1-|z_0|^2}\sqrt{1-|z_1|^2}z_2 & \cdots \\
 \sqrt{1-|z_0|^2}  & z_0^*z_1 & -z_0^*\sqrt{1-|z_1|^2}z_2 & \cdots\\
 0 & \sqrt{1-|z_1|^2} & z_1^*z_2 & \cdots \\
 0 & 0 & \sqrt{1-|z_2|^2} & \cdots \\
 \vdots & \vdots & \vdots &  \ddots
 \end{array}\right), \label{u}
\ee
which satisfies $\hat{\cal U}^\dag\hat{\cal U}=1$.

%%%%%%%%%%%%%%%%%%%%%%%%%%%%%%%%%%%%%%%%%%%%%%%%%%%%%%%%%%%%%%%%%%%%%%%%%%%%%%%
\subsection{One-dimensional quantum Ising model}

We apply the discrete-time formalism to the one-dimensional 
quantum Ising model in Eq.~(\ref{ising}).
It is well known that the model is integrable, and it can be written as a  free-fermion 
Hamiltonian, as described, e.g., in \cite{Sachdev11}.
The system is shown to be equivalent to the sum of independent two-level systems.
We assume $N$ is even and impose the periodic boundary condition $\hat{Z}_{N+1}=\hat{Z}_1$.
When the discrete-time is denoted as $t=-t_Q+k\Delta t$ with $k=0,1,\dots,M$, 
the final time is given by $t_Q=M\Delta t/2$. 
The effective Hamiltonian is written as 
\be
 \hat{H}^k=\sum_{n=1}^{N/2} \left(
 -\epsilon_n^k \hat{Z}^n+v_n^k \hat{X}^n\right),
\ee
where $\hat{Z}^n$ and $\hat{X}^n$ are Pauli operators at each site $n$, and 
\be
 && \epsilon_n^k=2h\sqrt{\left(1-\frac{2k}{M}\right)^2
 +4\left(1-\frac{k}{M}\right)\frac{k}{M}\sin^2\frac{p_n}{2}}, \\
 && v_n^k = -\frac{h^2\sin p_n}{t_Q(\epsilon_n^k)^2}, \\
 && p_n=\frac{\pi}{2N}(2n-1).
\ee
The initial state is set as $|\psi(0)\rangle=\otimes_n |0\rangle_n$.
At $k=M/2$ and $N\to\infty$, we find $\epsilon_n^k\to 0$,
which denotes the quantum phase transition for the corresponding static system.

We apply the unitary time-evolution operator in the Arnoldi iteration procedure at each step.
It is written as 
\be
 &&\hat{U}^k =\prod_{p=1}^{N/2}\left[\cos(h_n^k\Delta t)
 -i\left(\hat{Z}^n\frac{\epsilon_n^k}{h_n^k}
 +\hat{X}^n\frac{v_n^k}{h_n^k}\right)\sin(h_n^k\Delta t)\right],
\ee
where $h_n^k=[(\epsilon_n^k)^2+(v_n^k)^2]^{1/2}$.

%%%%%%%%%%%%%%%%%%%%%%%%%%%%%%%%%%%%%%%%%%%%%%%%%%%%%%%%%%%%%%%%%%%%%%%%%%%%%%%
\subsection{Full orthogonalization procedure}

The Arnoldi iteration procedure described in the main body involves large numerical errors
and we use the full orthogonalization~\cite{Rabinovici21}.
For a given initial basis $|K_0^k\rangle$ at each time step $k$ and 
a given set of the Krylov basis $|K_n^{k-1}\rangle$ ($n=0,1,\dots, k-1$) at $k-1$,
we produce the set at the next step $k$ as 
\be
 |K_n^k\rangle\sqrt{1-|z_{n-1}^{k-1}|^2} = \hat{U}^k|K_{n-1}^{k-1}\rangle
 -\sum_{m=0}^{n-1}|K_m^k\rangle\langle K_m^k|\hat{U}^k|K_{n-1}^{k-1}\rangle \qquad (n=1,2,\dots,k).
\ee
The time-evolved state $|\psi^k\rangle$ is calculated independently as 
$|\psi^k\rangle=\hat{U}^{k-1}|\psi^{k-1}\rangle$, 
and we take the overlap $\langle n|\varphi^k\rangle=\langle K_n^k|\psi^k\rangle$ 
to calculate the spread complexity.

%%%%%%%%%%%%%%%%%%%%%%%%%%%%%%%%%%%%%%%%%%%%%%%%%%%%%%%%%%%%%%%%%%%%%%%%%%%%%%%
\section{Krylov method for periodic systems}

%%%%%%%%%%%%%%%%%%%%%%%%%%%%%%%%%%%%%%%%%%%%%%%%%%%%%%%%%%%%%%%%%%%%%%%%%%%%%%%
\subsection{Time evolution in extended space}

We consider a periodic Hamiltonian satisfying $\hat{H}(t)=\hat{H}(t+T)$.
It can be decomposed as 
\be
 \hat{H}(t)=\sum_{m=-\infty}^\infty \hat{H}_m e^{-im\Omega t},
\ee
where $\Omega=2\pi/T$.
Although $|\psi(t)\rangle$ is not necessarily a periodic function, 
$|\psi(s,t)\rangle$ in Eq.~(\ref{tt'}) is periodic in $t$ if we choose $|\phi(t)\rangle$ 
as a periodic state, and we can write 
\be
 |\psi(s,t)\rangle = \sum_{m=-\infty}^\infty |\psi_m(s)\rangle e^{-im\Omega t}. \label{ft}
\ee
The initial condition $|\psi(0,t)\rangle=|\phi(t)\rangle$ gives us the relation 
\be
 |\phi(t)\rangle = \sum_{m=-\infty}^\infty |\psi_m(0)\rangle e^{-im\Omega t}.
\ee
Furthermore, setting $t=0$, we obtain
\be
 |\psi(0)\rangle = \sum_{m=-\infty}^\infty |\psi_m(0)\rangle.
\ee
A possible simplest choice is
\be
 |\psi_m(0)\rangle = |\psi(0)\rangle\delta_{m,0},  \label{initial}
\ee
which corresponds to choosing $|\phi(t)\rangle =|\psi(0)\rangle$.

By using the periodicity with respect to $t$, we consider the Fourier decomposition 
in Eq.~(\ref{ft}).
The transformed state satisfies 
\be
 i\partial_s|\psi_m(s)\rangle = \sum_{n=-\infty}^\infty
 (\hat{H}_{m-n}-\delta_{m,n}m\Omega)|\psi_n(s)\rangle.
\ee
According to the standard procedure, we extend the Hilbert space to 
the Hilbert--Floquet, or Sambe, space.
We define the state vector in the extended space as 
\be
 |\Psi(s)\rangle\!\rangle
 = \sum_m |\psi_m(s)\rangle\otimes |m\rangle
 =\left(\begin{array}{c}
 \vdots \\ |\psi_1(s)\rangle \\ |\psi_0(s)\rangle \\ |\psi_{-1}(s)\rangle \\ \vdots 
 \end{array}\right), 
\ee
and the inner product $\langle\!\langle \Psi'(s)|\Psi(s)\rangle\!\rangle=
\sum_m \langle\psi_m'(s)|\psi_m(s)\rangle$ is naturally defined from this representation.
The Schr\"odinger equation reads
\be
 i\partial_s|\Psi(s)\rangle\!\rangle = (\hat{\cal H}-\hat{\cal M}\Omega)|\Psi(s)\rangle\!\rangle,
\ee
where 
\be
 \hat{\cal H}-\hat{\cal M}\Omega = \left(\begin{array}{ccccc}
 \ddots & \ddots & \ddots && \\
 \ddots & \hat{H}_0-\Omega &\hat{H}_1 & \hat{H}_2 & \\
 \ddots & \hat{H}_{-1} & \hat{H}_0 & \hat{H}_1 & \ddots \\
 & \hat{H}_{-2} & \hat{H}_{-1} & \hat{H}_0+\Omega & \\
 &&\ddots & \ddots & \ddots
 \end{array}\right).
\ee
In this infinite Hamiltonian matrix, the $(m,n)$-block is given by 
$\hat{H}_{m-n}-\delta_{m,n}m\Omega$.
The initial condition corresponding to Eq.~(\ref{initial}) is given by 
\be
 |\Psi(0)\rangle\!\rangle=\left(\begin{array}{c}
 \vdots \\ 0 \\ |\psi(0)\rangle \\ 0 \\ \vdots \end{array}\right).
\ee

Since the effective Hamiltonian in the extended space is independent of $s$, 
we can apply the standard Krylov method for time-independent generators.
We set the initial basis as the initial state: 
\be
 |K_0\rangle\!\rangle = |\Psi(0)\rangle\!\rangle.
\ee
Then, higher-order basis elements are constructed through the recurrence relation 
\be
 |K_{k+1}\rangle\!\rangle b_{k+1}
 =(\hat{\cal H}-\hat{\cal M}\Omega)|K_{k}\rangle\!\rangle
 -|K_{k}\rangle\!\rangle a_{k}-|K_{k-1}\rangle\!\rangle b_{k},
\ee
where 
$a_k=\langle\!\langle K_k|(\hat{\cal H}-\hat{\cal M}\Omega)|K_{k}\rangle\!\rangle$ and 
$b_k=\langle\!\langle K_{k-1}|(\hat{\cal H}-\hat{\cal M}\Omega)|K_{k}\rangle\!\rangle$.

Suppose that we can find the Krylov basis and that the number of the basis elements, 
the Krylov dimension, is given by $d$.
Then, using the matrix 
$\hat{F}=(|K_0\rangle\!\rangle, |K_1\rangle\!\rangle, \dots, |K_{d-1}\rangle\!\rangle)$,
the effective Hamiltonian is transformed as 
\be
 (\hat{\cal H}-\hat{\cal M}\Omega)\hat{F}=\hat{F}\hat{\cal L},
\ee
where 
\be
 \hat{\cal L} = \left(\begin{array}{cccccc}
 a_0 & b_1 & 0 & &&\\
 b_1 & a_1 & b_2 & &&\\
 0 & b_2 & a_2 &&& \\
 &&&\ddots && \\
 && && a_{d-2} & b_{d-1} \\
 && && b_{d-1} & a_{d-1}
 \end{array}\right),
\ee
and the state is written as 
\be
 |\Psi(s)\rangle\!\rangle = \hat{F}|\Phi(s)\rangle = 
 \sum_{k=0}^{d-1}|K_k\rangle\!\rangle \Phi_k(s).
\ee
Note that $|\Phi(s)\rangle=\sum_{k=0}^{d-1}|k\rangle\Phi_k(s)$ follows 
the Schr\"odinger equation with the generator $\hat{\cal L}$ and is written as 
\be
 |\Phi(s)\rangle = e^{-is\hat{\cal L}}|0\rangle,
\ee
with the initial condition
\be
 |0\rangle = |\Phi(0)\rangle = \left(\begin{array}{c} 1 \\ 0 \\ \vdots \\ 0 \end{array}\right).
\ee

Now, we go back to the original Hilbert space. The state 
$|\psi(s,t)\rangle$ is generally written as 
\be
 |\psi(s,t)\rangle = \hat{\cal I}e^{-i\hat{\cal M}\Omega t}\hat{F} e^{-i\hat{\cal L}s}|0\rangle
 = \sum_{k=0}^{d-1}\hat{\cal I}e^{-i\hat{\cal M}\Omega t}
 |K_k\rangle\!\rangle\langle k| e^{-is\hat{\cal L}}|0\rangle,
\ee
where 
\be
 \hat{\cal I}=(\hat{I}_N, \hat{I}_N, \dots).
\ee
When the state is written in the standard Krylov form 
\be
 |\psi(s,t)\rangle = \sum_{k=0}^{d-1}|K_k(t)\rangle\Phi_k(s)
 = \sum_{k=0}^{d-1}|K_k(t)\rangle\langle k|e^{-is\hat{\cal L}}|0\rangle,
\ee
the basis $|K_k(t)\rangle$ is defined as 
\be
 |K_k(t)\rangle = \sum_m e^{-im\Omega t}\langle m|K_k\rangle\!\rangle, 
\ee
and satisfies the orthonormal relation 
\be
 \frac{1}{T}\int_0^{T} dt\,\langle K_k(t)|K_{l}(t)\rangle = \delta_{k,l}.
\ee
This condition corresponds to extending the Hilbert space.
States $|\psi\rangle e^{-im\Omega t}$ with $m=0,\pm 1,\dots$ are distinguished with each other.

%%%%%%%%%%%%%%%%%%%%%%%%%%%%%%%%%%%%%%%%%%%%%%%%%%%%%%%%%%%%%%%%%%%%%%%%%%%%%%%
\subsection{Lipkin--Meshkov--Glick model}

As an example, we consider the Hamiltonian in Eq.~(\ref{lmg}).
The effective Hamiltonian is given by 
\be
 \hat{\cal H}-\hat{\cal M}\Omega = \left(\begin{array}{ccccc}
 \ddots & \ddots & \ddots && \\
 \ddots & \hat{H}_0-\Omega & -i\hat{P} & 0 & \\
 \ddots & i\hat{P} & \hat{H}_0 & -i\hat{P} & \ddots \\
 & 0 & i\hat{P} & \hat{H}_0+\Omega & \\
 &&\ddots & \ddots & \ddots
 \end{array}\right),
\ee
where 
\be
 && \hat{H}_0=-\frac{2J}{N}(\hat{H})^2, \\
 && \hat{P}=h\hat{S}^x.
\ee
The explicit form of the Hamiltonian is denoted by using a fixed basis.
We use the $z$-basis set $\{|\mu\rangle\}_{\mu=0,1,\dots,N}$ satisfying 
\be
 \hat{S}^z|\mu\rangle=\left(\frac{N}{2}-\mu\right)|\mu\rangle.
\ee

When the initial state is given by $|0\rangle$, we choose the first Krylov basis as 
\be
 |K_0\rangle\!\rangle=|\mu=0\rangle\otimes|m=0\rangle.
\ee
We expand each basis as 
\be
 |K_k\rangle\!\rangle=\sum_{\mu=0}^N\sum_{m=-\infty}^\infty 
 |\mu\rangle\otimes|m\rangle f_k(\mu,m).
\ee
Then, the recurrence relation is written as 
\be
 b_{k+1}f_{k+1}(\mu,m) &=& 
 \left[-\frac{NJ}{2}\left(1-\frac{2\mu}{N}\right)^2-m\Omega-a_k\right]f_{k}(\mu,m)
 -b_kf_{k-1}(\mu,m) \no\\
 && -\frac{ih}{2}
 \sqrt{\mu(N+1-\mu)}\left[f_k(\mu-1,m+1)-f_k(\mu-1,m-1)\right] \no\\
 && -\frac{ih}{2}\sqrt{(\mu+1)(N-\mu)}\left[f_k(\mu+1,m+1)-f_k(\mu+1,m-1)\right].
\ee
The Lanczos coefficients are determined from the orthonormality condition
\be
 \sum_{\mu=0}^N\sum_{m=-\infty}^\infty f_k^*(\mu,m)f_l(\mu,m)=\delta_{k,l}.
\ee

When the modulated field is changed from $\sin\Omega t$ to $\cos\Omega t$ in Eq.~(\ref{lmg}),
the recurrence relation is slightly modified as
\be
 b_{k+1}f_{k+1}(\mu,m) &=& 
 \left[-\frac{NJ}{2}\left(1-\frac{2\mu}{N}\right)^2-m\Omega-a_k\right]f_{k}(\mu,m)
 -b_kf_{k-1}(\mu,m) \no\\
 && -\frac{h}{2}
 \sqrt{\mu(N+1-\mu)}\left[f_k(\mu-1,m+1)+f_k(\mu-1,m-1)\right] \no\\
 && -\frac{h}{2}\sqrt{(\mu+1)(N-\mu)}\left[f_k(\mu+1,m+1)+f_k(\mu+1,m-1)\right].
\ee
We confirm that the result of the Lanczos coefficients is basically unchanged.

\end{document}